\documentclass[fleqn,usenatbib,onecolumn]{mnras}

\usepackage{newtxtext,newtxmath}
\usepackage[T1]{fontenc}

\DeclareRobustCommand{\VAN}[3]{#2}
\let\VANthebibliography\thebibliography
\def\thebibliography{\DeclareRobustCommand{\VAN}[3]{##3}\VANthebibliography}

\usepackage{graphicx, xcolor, subcaption}	% Including figure files
\usepackage{amsmath}	% Advanced maths commands

\newcommand{\br}{\mathbf{r}}
\newcommand{\boldv}{\mathbf{v}}
\newcommand{\boldB}{\mathbf{B}}
\newcommand{\boldxi}{\boldsymbol{\xi}}
\newcommand{\boldk}{\boldsymbol{k}}
\newcommand{\boldG}{\mathbf{G}}

\newcommand{\omegat}{\tilde{\omega}}

\newcommand{\real}{\text{Re}}

\usepackage{soul}

\title[Visualising accretion disk instabilities]{A visual approach to global accretion disk instabilities}

\author[N. Brughmans \& R. Keppens]{
Nicolas Brughmans,$^{1}$\thanks{E-mail: nicolas.brughmans@kuleuven.be}
Rony Keppens,$^{1}$
\\
$^{1}$Centre for mathematical Plasma-Astrophysics, KU Leuven, Celestijnenlaan 200B, 3001 Leuven, Belgium.\\
}

\date{Accepted XXX. Received YYY; in original form ZZZ}

\pubyear{2024}

\begin{document}
\label{firstpage}
\pagerange{\pageref{firstpage}--\pageref{lastpage}}
\maketitle

\begin{abstract}
    For over 30 years, the Magneto-Rotational Instability has been accepted as the mechanism driving accretion disk turbulence. Its physical basis is well understood, where an interplay between centrifugal forces and magnetic tension transfers angular momentum between oppositely displaced fluid elements. In this work, we revisit this picture in global disk models and various magnetic field topologies and generalise it to non-axisymmetric instabilities like the Super-Alfvénic Rotational Instability (SARI). We use the open-source \texttt{Legolas} software to quantify all complex-valued linear eigenfunctions for the (near-)eigenmodes and visualise the resulting spatio-temporal variations in real space in a manner that can be compared to direct numerical simulations of disks. The field perturbations are fundamentally different between the (axisymmetric) MRI and the novel, ultra-localised SARI modes, which bear some resemblance to spiral modes in galaxies but differ in important ways. We use a combined numerical-analytical approach to study the polarization of the magnetic and velocity field perturbations. Finally, we compare disks of differing magnetic topology where many linear modes are merely superposed to recreate a visual impression of `turbulent' fields and quantify the resulting stresses. We find that even superposed, still linearly growing SARI  modes can already provide the needed effective viscosity-related $\alpha$ values invoked for angular momentum  transport. 3D views on the magnetic field perturbations show that SARI modes of opposite azimuthal mode number $m$ may naturally introduce plasmoid and toroidal flux-tube like field deformations.
\end{abstract}


\section{Introduction}

Accretion disks are theorised to be highly unstable to magnetohydrodynamic (MHD) instabilities like the magneto-rotational instability (MRI) \citep{BH98}. Considerable work has been done over the last 30 years to show that weakly magnetised, global disks, as well as local shearing boxes indeed feature turbulent dynamics which has routinely been interpreted as a result of the linear, axisymmetric MRI, with resulting Maxwell and Reynolds stresses leading to an effective viscosity analogous to the Shakura-Sunyaev model \citep{Hawley95,bai_stone,bacchini2022,bacchini2024}. Especially for shearing boxes, much attention has been devoted to the so-called \textit{channel mode} solution that is then attributed to the linear MRI perturbation in a box, which magically also satisfies the non-linear shearing sheet equations and can itself be unstable to secondary parasitic instabilities that are non-axisymmetric \citep{goodmanxu94, latter09}. At the same time, global disks have also been shown to produce turbulence as a result of the MRI \citep{salvesen16, mishra22, ripperda22, fragile23}, but direct numerical studies of (non-linear) global disk evolutions must of course compromise on achievable resolutions. In this paper, we will instead use exact linear eigenoscillations from the spectrum of all MHD eigenmodes of a specific disk model, where we do not need to make any resolution compromise.

The basic physical mechanism of the MRI is the same in all these systems, with a delicate interplay between magnetic field tension and centrifugal/Coriolis forces driving the instability. This basic `cartoon' of the axisymmetric MRI is well-established and accepted as the main driver of accretion disk turbulence. However, insights into non-axisymmetric MRI counterparts remain scarce and divided: in local shearing sheet approaches \citep{BH92b}, the instability is transient because its radial wavenumber is sheared to larger values and eventually leads to stability; in global approaches, the nearness of `walls' (through enforced radial boundary conditions) is cumbersome but the instabilities are true, growing normal modes \citep{ogilvie_pringle}. Nevertheless, a renaissance of the topic might be on its way: (1) in recent years, globally treated discrete non-axisymmetric modes have been shown to open up more field strengths and topologies to instability \citep{begelman22,begelman23,BKG24}. Such arguments provide strong indications that e.g. for strong vertical fields, where the usual MRI is stabilised, these discrete non-axisymmetric modes may play a large role in providing instability; (2) direct observations in laboratory plasmas \citep{seilmayer14, wang2022non-axisymm,observation2024}, where modes manifesting themselves as discrete growing modes near actual walls play a large role. Finally, most relevant for this paper: (3) the existence of non-axisymmetric modes that are radially and vertically localised, insensitive to radial boundary conditions, but global in nature in the sense that one adopts a non-zero azimuthal mode number in the analysis, which can however be taken large to represent locality in azimuth. These modes are for all intents and purposes true normal Fourier modes (growing exponentially!): the Super-Alfvénic Rotational Instability (SARI) \citep{GK22,BKG24}, which come in both discrete (i.e. exact eigenmodes with specific complex eigenfrequencies) as well as quasi-continuum variants (i.e. near-eigenmodes that occupy entire 2D regions in the complex eigenfrequency plane). Because of the novelty that these newly discovered modes are insensitive to the walls, which are usually an unwanted, unphysical factor in astrophysical disks, this paper will focus on such quasi-continuum SARIs. Technically, they require a minute (e.g. below machine precision) addition of energy to the system, which is de facto satisfied for all direct numerical simulations. This energy addition is not needed for the discrete SARIs, as studied by e.g. \cite{ogilvie_pringle}, which are equally important for a theoretical understanding of non-axisymmetric modes, and might now very well have been observed in the laboratory. However, a physical and intuitive picture of non-axisymmetric modes, similar to the classical cartoon of the MRI, is still missing. 
% This contrasts with the state of knowledge in galactic disks, where much of the original research has focused on large-scale spiral patterns in galaxies \citep{lin_shu64, toomre69,lynden-bell_ostriker}. 

In this paper, we aim to gain an improved understanding of these novel non-axisymmetric modes as compared to the axisymmetric MRI through visualisation of their normal Fourier mode structure in space and time, combined with analytical efforts to back up these findings. This effort is non-trivial because the eigenfunctions involved are complex-valued functions that situate in Fourier space, and hence some of the physical appearance of the modes in real 3D and time lies hidden. We first set the stage for a linear MHD stability analysis of a cylindrical disk in Sec.~\ref{sec:linear_mhd}. We then use these tools to recapitulate the most important properties of the non-axisymmetric SARIs, and highlight implications of their non-axisymmetry, notably spiral structure and non-zero phase speeds, in Sec.~\ref{sec:SARI_spirals}. We then revisit the standard MRI cartoon in Sec.~\ref{sec:field_topology}, which is shown to be captured by this global model, and then generalise this picture to include SARIs and fields with an azimuthal component. In Sec.~\ref{sec:many_modes} we look at the implications that many superposed MRI or SARI modes have on the global appearance of disks, as well as on the averaged stresses and the corresponding Shakura-Sunyaev $\alpha$-parameter. Finally, in Sec.~\ref{sec:local_vs_global}, we connect the SARIs to earlier results on localised non-axisymmetric instabilities. Appendix~\ref{sec:efs_analytical} analytically confirms the polarization properties of the modes discussed in Sec.~\ref{sec:field_topology}.

\section{Linear ideal MHD} \label{sec:linear_mhd}

\subsection{Normal mode analysis of a cylindrical disk} \label{sec:equilibrium}

A global, thin disk can be represented by a cylindrical annulus of plasma rotating about a central object, where vertical variation is ignored by assuming modes to be vertically localised around the midplane. In its most basic form, such a steady-state, ideal disk can be described by self-similar profiles for the MHD quantities \citep{spruit1987}, with a general equilibrium magnetic field having a vertical and azimuthal component. Since we are dealing with ideal MHD, it is expedient to use non-dimensionalised quantities such that $r\in[1, 2]$, $\rho(1)=1$ and velocities are given in terms of the usual Keplerian rotation velocity $v_{K1}=\sqrt{{GM_*}/{r_1}}$, where $r_1$ is the inner disk radius, serving as length unit. The requirement that modes be vertically localised can then be written in terms of the axial mode number as $k \gg 1$.  As in our previous work, we use the following profiles:
\begin{equation} \label{eq:equilibria}
    \rho_0(r) = r^{-3/2}, \quad p_0(r) = p_1 r^{-5/2}, \quad \boldv_0(r) = \Omega_1 r^{-1/2}\boldsymbol{\hat{\theta}}, \quad \boldB_0(r) = r^{-5/4} \left(B_{\theta1}  \boldsymbol{\hat{\theta}} + B_{z1}\mathbf{\hat{z}}\right) = B_{\theta 0}(r) \boldsymbol{\hat{\theta}} + B_{z0}(r) \mathbf{\hat{z}},
\end{equation}
which are fully determined by the three parameters
\begin{equation} \label{eq:parameters}
    \beta = 2p_1/B_1^2, \quad B_{z1}, \quad B_{\theta1}
\end{equation}
that denote values at the inner boundary \citep{GK22, BKG24}. For the current exposition, it suffices to note that in a true, explicit balance between forces in the radial direction, the rotation frequency $\Omega_1$ is modified from its usual Keplerian value of $1$ by pressure and Lorentz forces. Nevertheless, for the application to a thin disk in this work, $\Omega_1 \approx 1$.

Since the original MRI \citep{BH98} and SARI \citep{GK22,BKG24} analysis applies to thin disks, we will only consider weakly-magnetised disks with $B_1 = 0.01$, $\beta = 10$. The orientation of the field is controlled by the inverse pitch angle $\mu_1 = B_{\theta1}/B_{z1}$, and we consider vertical ($\mu_1 = 0$), helical ($\mu_1 = 1$) and nearly azimuthal ($\mu_1 = 10$) background fields in this work. In the thin disk regime, the rotation is super-Alfvénic (i.e. faster than the Alfvén speed $v_A = B/\sqrt{\rho}$): if $p_1 \ll 1$, $B_1 \ll 1$, it follows automatically that $v_A \ll 1 \approx \Omega_1$. Note that the true inverse pitch angle $\mu(r) = B_{\theta0}(r) / rB_{z0}(r)$ usually varies over the domain. In this work, we always take $r\in[1,2]$. The ratio of specific heats is set to $\gamma = 5/3$. It was clarified in our earlier papers that the MHD eigenmode analysis can be done for even more general disks, also strongly magnetised ones, and the numerical tools we use already confirmed that our analytical SARI results are not restricted to weakly magnetised thin disks. However, the main point of this paper is to contrast the MRI with the novel SARI modes, and this is therefore done in the original weak-field setting.

Assuming the usual Fourier definition of perturbations with azimuthal wavenumber $m$, vertical wavenumber $k$, and frequency/growth rate $\omega = \omega_r +i\nu$ for a normal mode in such an equilibrium,\begin{equation} \label{eq:fourier}
    f(r,\theta,z,t) = \hat{f}(r) \exp{\left[i\left(m\theta+kz - \omega t\right)\right]},
\end{equation}
it is possible to transform the MHD equations into two different kinds of eigenvalue problems (see Sect.~\ref{sec:mhd_spectroscopy}). Here, a plane wave decomposition is taken in the trivial azimuthal and vertical directions, whereas the non-trivial radial structure of the eigenmodes is captured by the Fourier coefficients (or amplitude eigenfunctions), which are functions of radius. The SARI analysis assumed $k^2r^2 \gg m^2$ such that these modes are vertically localised. For consistency in the cylindrical model, we should exclude values of $k$ for which this assumption doesn't hold. In general, the MRI (and by extension SARI) is assumed to be stabilised whenever the vertical wavelength $L = 2\pi / k$ exceeds the vertical size of the disk, i.e. $L \leq 2H(r)$ where $H = \sqrt{p_1}r$ is the pressure scale height of the disk \citep{Hawley95}. In our equilibrium, this implies that values of $k > 70 / r$ are consistent. 

\subsection{MHD spectroscopy} \label{sec:mhd_spectroscopy}
The linear dynamics of an MHD equilibrium can be described in two ways: (1) the equation of motion for the Lagrangian displacement of a fluid element \citep{frieman-rotenberg1960}, which for the case of a cylindrical accretion disk can be reduced to a second-order ODE for the radial displacement $\xi$ \citep{keppens2002}, or (2) a generalised matrix eigenvalue problem directly obtained from linearising the full (Eulerian) MHD equations. The latter is used routinely for computing MHD eigenmodes in cylinders or plane-parallel atmospheres \citep{claes2020legolas} and tokamaks or accretion tori \citep{castor1998,phoenix2007}. These two approaches are equivalent means to quantify the full MHD spectrum of natural eigenmodes for a given background equilibrium (e.g. a magnetised accretion disk), as a field-theoretical treatment of the time-reversible ideal MHD equations also exposes the same set of self-adjoint operators that govern the linear dynamics at any time during a non-linear MHD evolution \citep{keppens2016}.

For a given, general equilibrium, a modern open-source tool like \texttt{LEGOLAS}\footnote{\url{http://legolas.science}} can determine the full Fourier spectrum of waves and instabilities by using well-known linear algebra tools \citep{claes2020legolas,legolasjordi,legolas2}, concurrently solving for the eigenvalues and eigenfunctions of the perturbed MHD quantities. In ideal MHD, such an MHD spectrum contains the eigenfrequencies of all possible modes for a given wavevector $(m/r, k)$, including the usual slow, Alfvén and fast modes. Furthermore, because the operators involved are self-adjoint, these spectra are always up-down symmetric, implying that an instability with growth rate $\nu$ has a corresponding damped version with damping rate $-\nu$. Finally, these spectra for essentially one-dimensionally varying equilibria like disks and cylinders are organised by the continuous ranges of Alfvén and slow frequencies, to which the discrete eigenvalues cluster. A complete description of MHD spectroscopy, from static to stationary MHD in Cartesian and cylindrical systems can be found in \citet{GKP19}. MHD spectroscopy of cylindrical plasmas with equilibrium flow has not been used as extensively as in laboratory plasmas, but there are some recent examples for solar coronal flux tubes or loops \citep{skirvin24, joris24}.

The \texttt{Legolas} code was already applied to instabilities in accretion disks in \citet{BKG24}, where we confirmed and augmented the predictions on non-axisymmetric SARI modes of both the discrete and the quasi-continuum type. \texttt{LEGOLAS 2.0} \citep{legolas2} now supports visualisation of the eigenmodes to investigate the linear eigenmode behaviour in time and space. This is done by straightforward evaluating of the normal mode definition \eqref{eq:fourier} and letting the temporal and/or spatial coordinates vary to obtain the full 3D solutions. We are hence showing how one or many linear perturbations evolve, but all of these modes are exact Fourier solutions. We can use the freedom in amplitude and phase for each linear eigenmode to show what linear superpositions of exact modes look like, in which case they will behave independently of each other. We first evaluate the full, complex-valued, solutions and then plot the real part of the `wavefunction', in analogy with quantum mechanics:
\begin{equation} \label{eq:fourier_visual}
    f(\br,t) = \text{Re} \sum_j c_j \hat{f}_j(r) \exp{\left[i\left(m_j\theta+k_jz - \omega_j t\right)\right]},
\end{equation} 
where the index $j$ runs over the different superposed modes and $c_j$ is an arbitrary complex factor that represents both an amplitude and a rotation of the (complex) eigenfunctions. 

Because the theoretical tools for linear MHD have been developed using mostly the Lagrangian viewpoint, it is useful to list some conversions between the Eulerian and Lagrangian viewpoints for later use. The Eulerian perturbations can be related to the Lagrangian perturbations in the following way:
\begin{equation} \label{eq:eulerian_lagrangian}
    f_{1L} = f_{1E} + \boldxi\cdot\nabla f_0,
\end{equation}
where all these quantities are functions of $(\br, t)$. In particular, then, for a stationary equilibrium, the perturbed velocity is given by
\begin{equation} \label{eq:velocity_xi}
    \boldv = \boldv_0 + \boldv_0\cdot\nabla\boldxi + \frac{\partial\boldxi}{\partial t} - \boldxi\cdot\nabla \boldv_0 
    \qquad \Rightarrow \qquad 
    \boldv = -i\omegat \boldxi - r\Omega' \xi_r \boldsymbol{\hat{\theta}}
\end{equation}
for our assumed equilibrium. Here, $\omegat = \omega - m\Omega(r)$ is the wave frequency in the comoving frame, where $\Omega(r)=v_{\theta0}(r)/r$.

In accord with the usual Frieman-Rotenberg general formalism \citep{frieman-rotenberg1960}, the Lagrangian displacement vector $\boldxi$ satisfies the spectral equation
\begin{equation} \label{eq:spectral_equation}
    \mathbf{G}(\boldxi) - 2\rho\omegat U\boldxi + \rho\omega^2\boldxi = 0,
\end{equation}
where $\mathbf{G}$ is a generalised force operator, extending the static MHD force operator \citep{Bernstein1958} to include Doppler-Coriolis effects of rotation (and gravity), and $U$ is the Doppler shift operator. Both $U$ and $\mathbf{G}$ are self-adjoint, which can be exploited to locate eigenfrequencies in the complex $\omega$-plane through the spectral web-approach, which uses a shooting method involving the potential energy and Doppler shift \citep{goedbloed2018spectralI,goedbloed2018spectralII}. The difficulty of solving Eq.~\eqref{eq:spectral_equation} hence lies in it being a quadratic equation in $\omega$, leading to truly complex-valued solutions (as opposed to exclusively real $\omega^2$ values when analysing static ideal MHD equilibria). 

\subsection{Linear dynamics} \label{sec:defining_linear}

Using Eq.~(\ref{eq:fourier_visual}), we could in principle visualise the ensuing dynamics for arbitrary times as superposed on our equilibrium disk, except that we must exclude unphysical evolutions (like negative total densities, pressures or temperatures), and that eventually our linear analysis will break down. When looking at 3D mode structure, the question of what we mean by `linear' quickly arises: even though the linear dynamics may leave an imprint on the eventual non-linear dynamics (as in the channel flow solution, which is an actual non-linear solution of the shearing sheet equations), it is a priori unclear how fast non-linear effects take over after the linear stage. In the derivation of the linearised MHD equations, one assumes that perturbations are `small' with respect to the background, i.e. $|f_1| \ll |f_0|$. This definition is not unambiguous since it does not apply to perturbed quantities that do not play a role in the equilibrium (in our application, $v_{r0} = v_{z0} = 0$, for example). In static MHD, this question is even more pressing, given the absence of any background flow. 

We define `linear' as the phase where the perturbed magnetic field is everywhere weaker than the equilibrium field $B_0$. Put otherwise, 
\begin{equation} \label{eq:criterion_linear}
    B_r \lesssim B_0, \qquad B_{\theta,z} \lesssim B_{\theta0,z0},
\end{equation}
over the entire domain. If the equilibrium field is purely vertical/azimuthal, the complementary perpendicular component should also be compared with $B_0$. This proves to be the most stringent condition because the magnetic field is affected the most by the instability as a result of the mostly incompressible nature of the MRI/SARI in weakly magnetised disks, and is consistent with earlier studies \citep{Latter2015}. We check this condition at $\theta=0, z=0$, which allows us to use the eigenfunctions directly instead of comparing the full 3D variation over the disk. The absolute values of $B_{r,\theta,z}$ (which can be complex functions) are compared to the equilibrium background locally. Given the exponential growth $B_r \sim e^{\nu t}$ from complex eigenfrequency $\omega=\omega_r+i\nu$, criterion \eqref{eq:criterion_linear} limits the times that can be used for visualisation. We always compare the total perturbation of all superposed modes to the background locally. Because the maximal growth of the MRI/SARI happens at a rate of $\sim 0.75 \Omega$, the linear phase as defined here only lasts on the order of several rotation periods. We are hence looking at the short-time-scale behaviour of these modes. A second aspect of `linear' is that the perturbation has to start off small, too. Again for consistency with \citet{Latter2015}, we initialise modes with $B_r = 10^{-4} B_0$ unless mentioned otherwise. 
%We take this convention into account when showing single modes, and for visualisations containing many modes, we additionally randomize the mode amplitudes with a factor $10^{-4} - 10^{-2}$, and again initiate the maximum total perturbation at the same value.

\section{SARIs as spiral modes}  \label{sec:SARI_spirals}

\subsection{SARIs in thin disks: recapitulation}
Earlier work described non-axisymmetric modes of a new variety, termed quasi-mode SARIs \citep[][from hereon GK22]{GK22}. These modes were confirmed by \citet{BKG24} to be indistinguishable from proper normal modes by any numerical code, because they can be excited with very little added energy (below machine precision). The SARIs have several main properties:
\begin{enumerate}
    \item They make up a 2D region in the complex frequency plane that reaches to significant growth rates, essentially as big or sometimes exceeding the growth for the discrete axisymmetric MRI modes.
    \item This region is located above the overlapping forward and backward Alfvén continua, which implies that the eigenfunctions have two resonances with either continua. These resonances localise the eigenfunctions away from the boundaries, making them insensitive to the boundary conditions. In essence, these modes show surface (skin) current variations at these resonances, but these are dwarfed by their finite wave-package variations concentrated at the corotation radius.
    \item The eigenfunctions are centered around their corresponding corotation radius where $\omega_r = m\Omega(r_*)$, which is not a singularity in MHD, and are stationary in the comoving frame. This means that the disk can be full of localised wave packages around certain radii and comoving with the super-Alfvénic flow.
    \item These modes obey the same instability criterion as the MRI, where the Alfvén frequency $\omega_A = mB_{\theta0}/r + kB_{z0}$ needs to satisfy $0 < \omega_A(r) < \sqrt{3} \Omega(r)$ locally.
\end{enumerate}

Furthermore, there are discrete non-axisymmetric modes that were also analysed by GK22, but these modes are sensitive to one of the wall boundaries. In this work, we therefore only consider SARIs of the quasi-continuum variety, but most of the results on mode structure also apply to the discrete modes. We compare these SARIs to the global MRI, which significantly differs from non-axisymmetric modes in that it has no resonances (these depend on a non-zero Doppler shift $m\Omega$ of the frequencies) and is truly global because it is connected to both boundaries. For illustrative purposes, we will consider a quasi-continuum SARI that is fairly global in this Section, but it should be noted that the separation of the resonances depends on many factors (e.g. it increases at larger $r$ but decreases for larger $B_{\theta0}$ or $m$). Hence, the results shown here are also valid for truly localised SARIs. Examples of these latter, ultra-radially-localised quasi-continuum SARIs were already shown in Figure 11 of our earlier work \citep{BKG24}, and predicted to occur routinely in actual radially extended disks.

\subsection{Spiral shape}
\begin{figure}
    \centering
    \begin{subfigure}{0.59\columnwidth}
        \includegraphics[width=0.99\linewidth]{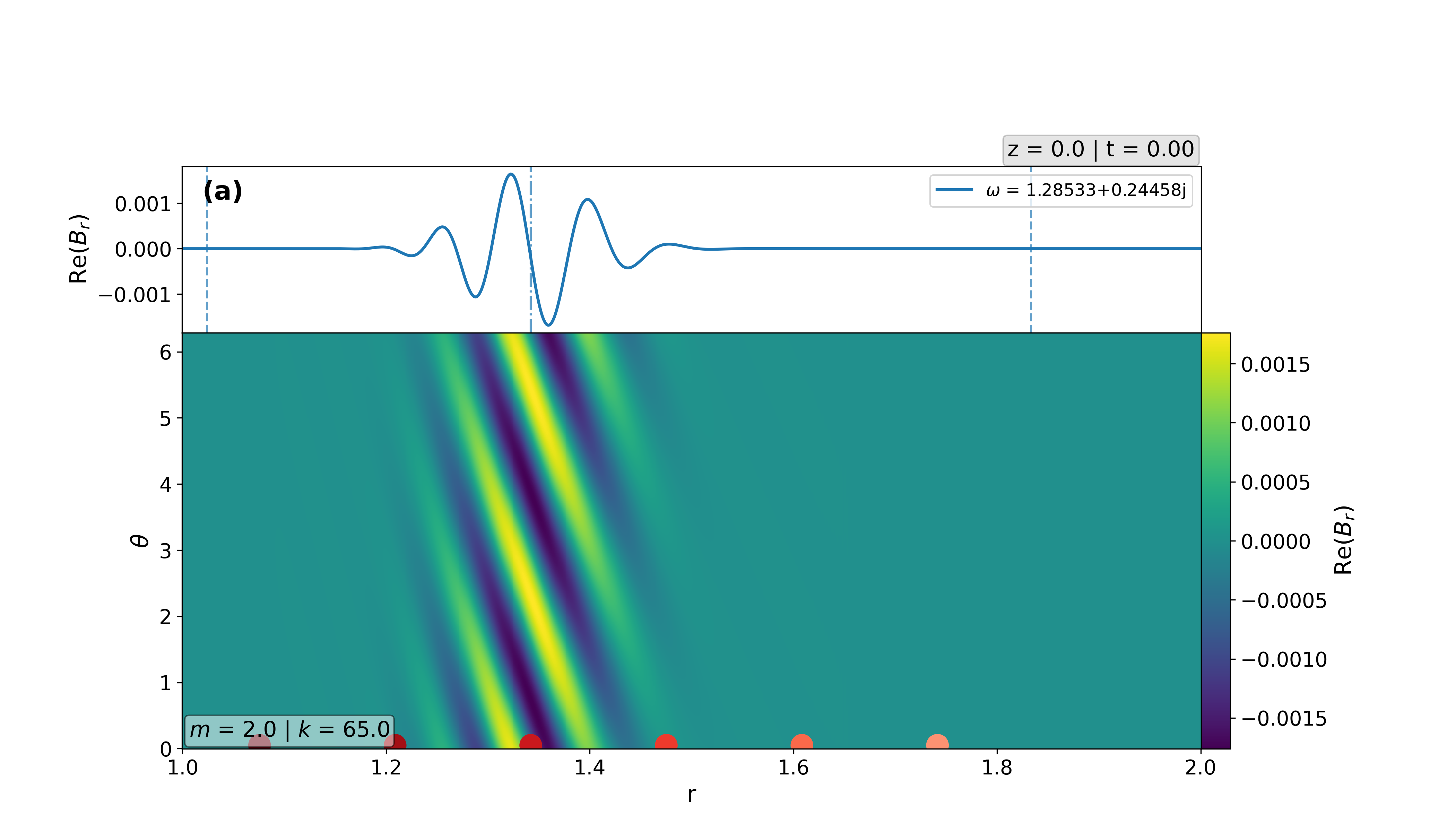}
    \end{subfigure}
    \begin{subfigure}{0.39\columnwidth}
        \includegraphics[width=0.99\linewidth]{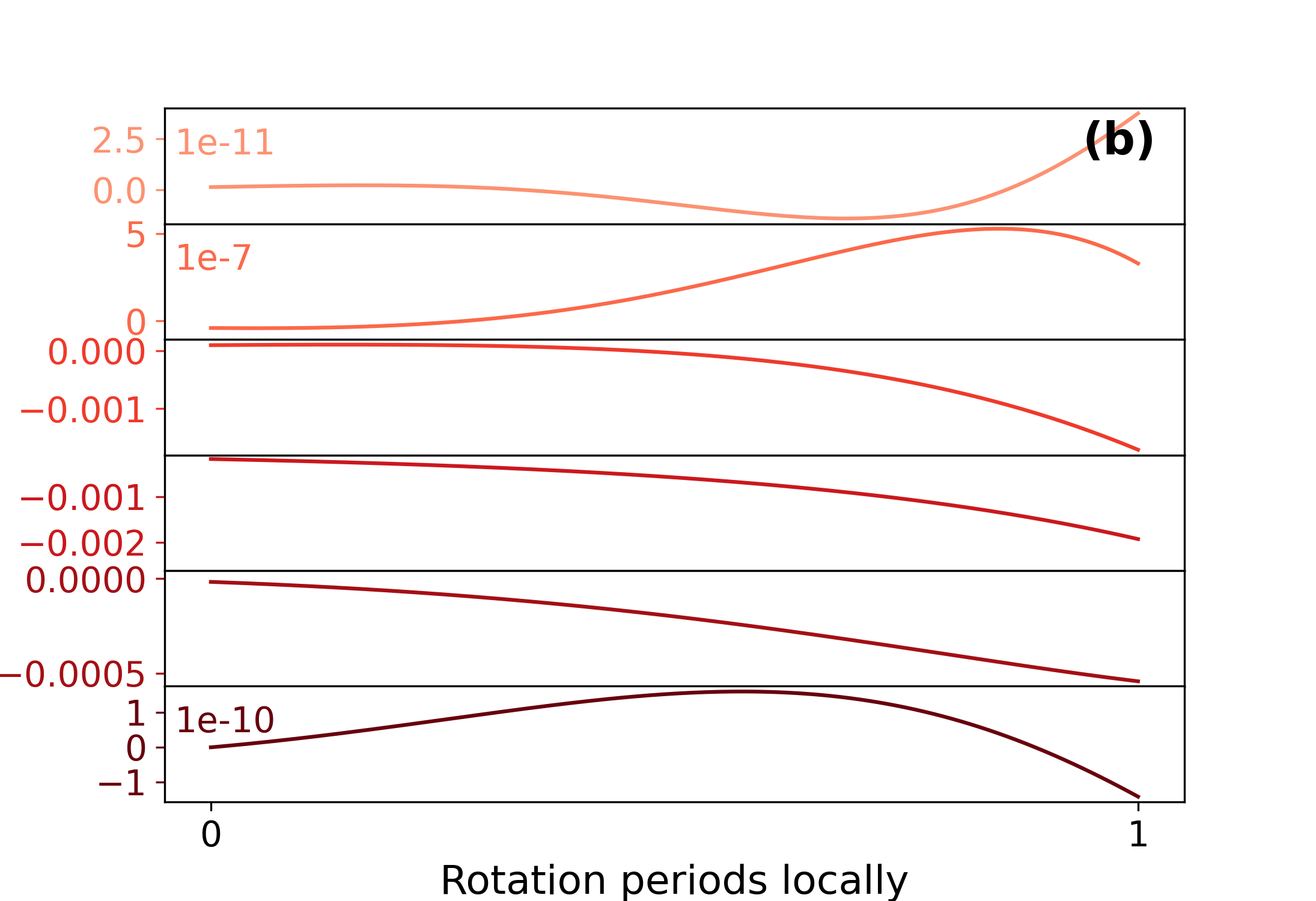}
    \end{subfigure}
    \caption{(a) Mode structure of a SARI ($m=2$, $k=65$, $\mu_1=0$) with eigenfrequency $1.28533+0.24458i$. Snapshot of $B_r$ at $t=0$ of a horizontal cut at $z=0$. The number of spiral arms corresponds to the value of $|m|$. (b) Eulerian perturbations of $B_r$ encountered by the red `corks' at different radii moving with the equilibrium flow at the bottom of Panel (a). Even at the Alfvén resonances far away from corotation, where the perturbation almost vanishes, the oscillatory growth is negligible and the mode hence appears stationary. An animated version of this figure is available in the online version of this paper, which then quantifies how this equatorial view on this $m=2$ SARI mode structure evolves in time as compared to the background flow.}
    \label{fig:spiral}
\end{figure}

Figure~\ref{fig:spiral}~(a) shows an $r-\theta$ cut of the perturbed $B_r$ component of the magnetic field for an $m=2$ SARI at time $t=0$ in a vertical equilibrium field, i.e. the background field is oriented perpendicular to the (equatorial) plane shown here. The resonance locations mentioned above, as well as the corotation radius of this mode, have been indicated on the figure, as they are shown as vertical (dotted and dash-dotted) lines in the top panel that shows the real part of $B_r(r)$. The equatorial plane view shows that the mode follows a trailing spiral, with the number of spiral `arms' equal to the wavenumber $|m|=2$ in this case. This is a well-known fact that applies to non-axisymmetric instabilities in general \citep{lynden-bell_ostriker,papaloizou-pringle, BH98}. These spiral arms propagate with the rotation frequency of the peak of the wave package, meaning that their shape is conserved. This behaviour is in stark contrast with the transient non-axisymmetric MRI in shearing box studies, where the mode is `stretched' by the azimuthal shear, leading to an increasing radial wavenumber that quenches the instability. However, it must be understood that the shearing box approximation in fact evaluates and views dynamics in a special, locally corotating frame which neglects global curvature effects, and where our $\theta$ direction becomes a periodic $y$ coordinate (i.e. assumes a high $m$), and the `radial' direction behaves as a time-shifted periodic Cartesian one. This assumption of local-corotation implies a non-trivial transformation between the lab-frame eigenfrequency-eigenvector pair $\omega-\boldxi$, the Doppler range $\tilde{\omega}(r)$ which quantifies all local comoving frequencies, and frequencies as evaluated in this non-inertial, accelerated frame with corresponding artificial forces. From Figure~\ref{fig:spiral} it is clear that the $m=2$ quasi-continuum SARI mode is not a transient structure, but an exponentially growing instability. Because of its radial localisation, the wave forms a stationary structure as seen from a comoving frame located at the corotation radius (dashdot blue line). A true polar projection of the mode structure reveals that the spirals are in fact tightly wound, especially for the more localised higher-$m$ modes. Figure~\ref{fig:spiral}~(b) shows the perturbation with respect to the equilibrium state as experienced along five red dots indicated in Fig.~\ref{fig:spiral}~(a) which move along with the background flow (see the online animation): this hence quantifies the local comoving behavior of the mode. The wave indeed appears almost comoving near the Doppler radius (see the third row from the bottom of panel (b)) since the growth is nearly purely exponential. Further away from corotation, the mode shows some appreciable oscillation occurring over one rotation period, which is about the time it takes the instability to grow to a non-linear amplitude. It is hence unclear how strong the spiral shape of these modes will be imprinted on the longer-time-scale dynamics of the system, as these linear modes are inherently short-lasting. The phase speeds associated with these oscillations will be the topic of Sec.~\ref{sec:phase_speeds}. The phase shift will be more noticeable at lower wavenumbers $m$, where the mode is more global. Note that at radii further away from corotation, the linear phase of one SARI lasts longer than at corotation because the perturbation is smaller. Hence, the number of oscillations before the mode degenerates non-linearly will be larger than that shown on Fig.~\ref{fig:spiral}, and the phase speeds will be noticeable. For larger $m$, the SARI is much more localised and will hence be close to stationary over the entire radial extent between its resonances.

These observations warrant some further discussion. First, the fact that the spiral shape is stationary in the corotating frame is a natural consequence of the structure of the eigenvalue problem: GK22 showed through a quadratic form (Eq.~A10) that any unstable mode necessarily corotates ($\omega_r = m\Omega(r_*)$ for some $r_*$), which is generally true in MHD \citep{acheson1973}. This actually implies that both the overstable SARI and its damped variant are corotating, since the pattern speed $\omega_r/m$ is independent of the sign of $m$. This is in correspondence with earlier results for non-uniform background fields. For solid body rotation, however, instabilities would propagate against the flow \citep{acheson_hide1973}. Second, the appearance of a spiral shape is in fact related to an MHD variant of the anti-spiral theorem \citep{lynden-bell_ostriker} in self-gravitating galactic disks, which states that a spiral-shaped perturbation can only be produced by an instability, and not by a propagating, non-degenerate, pure wave. Originally intended to disprove the idea that large-scale spirals in galaxies are steady oscillating waves, the theorem argues that spirals always appear in two flavours: one trailing and one leading \citep{toomre_review}. Because of the similar underlying structure of the eigenvalue problem, an analogous statement can be made for the SARIs as compared to their damped counterparts and to the purely oscillating non-axisymmetric Alfvén waves, see Sec.~\ref{sec:phase_speeds}. Third, there is virtually no difference between the spiral structures of $\pm m$ modes, and both propagate along with the flow. In that sense, the nomenclature `counter- and co-propagating' SARIs which was introduced in \cite{GK22} for the purely discrete SARI branches is misleading because either mode is comoving with the flow in the laboratory frame, but the wavefronts are slanted in the direction of the flow or against it (at constant $r$, the contours of constant phase depends on the sign of $m/k$). This implies that for the phase speeds, the distinction becomes more clear.

GK22 interpreted the localisation of the SARI as a result of wave pressure from forward and backward Alfvén waves emanating from the resonances, which are generally at some finite distance away from the corotation locus, interfering constructively to produce a wave package around corotation. Indeed, at marginal stability, the Alfvén waves inside and outside corotation might be viewed as negative and positive energy density waves, respectively, combining to form an instability \citep{haverkort}. Further study is needed to determine in how far parallels might be drawn with the Papaloizou-Pringle instability or the situation in galactic disks, where waves reflected off Lindblad resonances transfer energy at the corotation \textit{singularity} \citep{BH98}. 

\subsection{Phase speeds} \label{sec:phase_speeds}
\begin{figure}
    \centering
    \includegraphics[width=0.65\columnwidth]{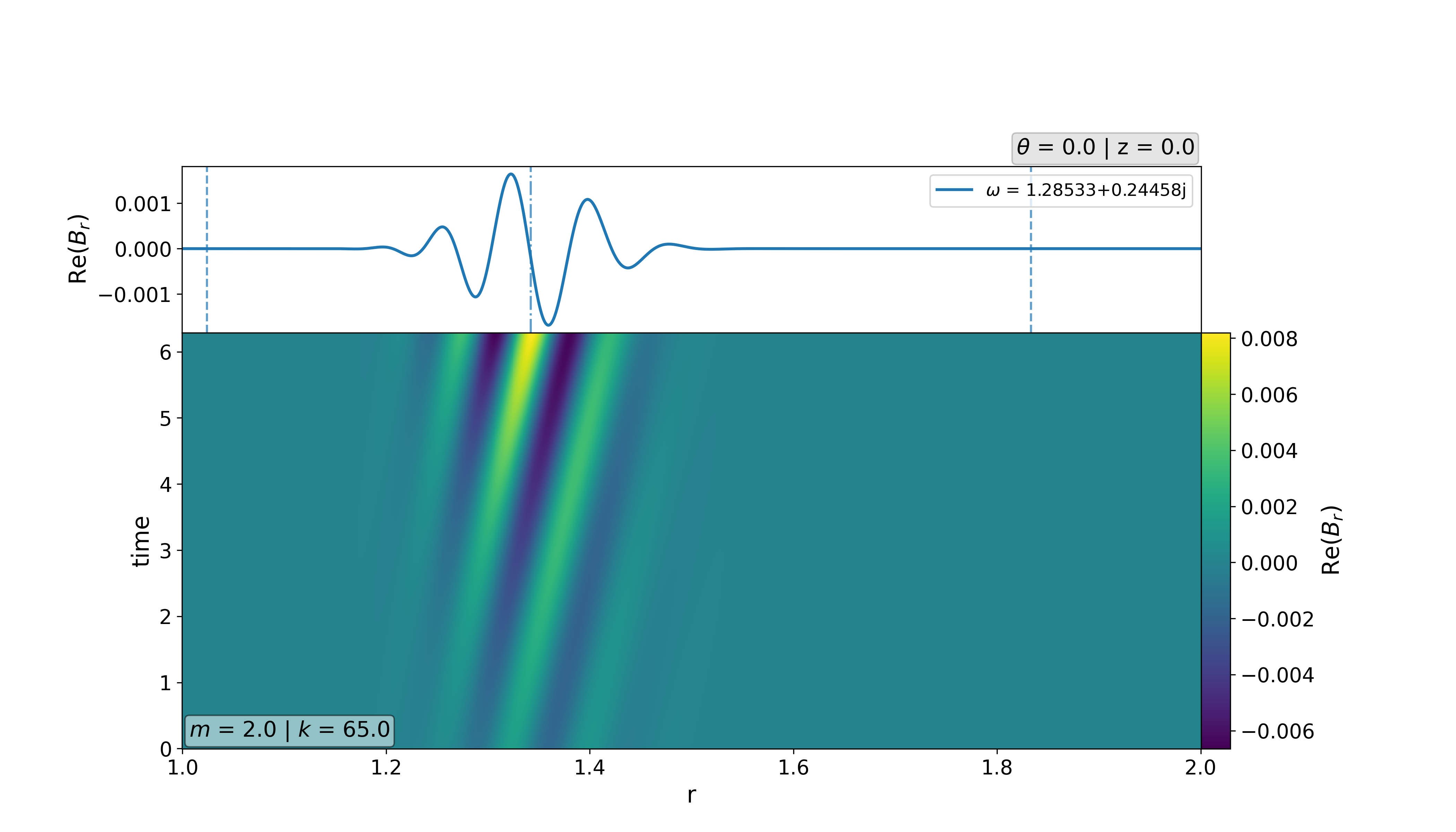}
    \caption{Time evolution of a SARI ($m=2$, $k=65$, $\mu_1=0$, $\omega=1.28533+0.24458i$) in a radial slice at $\theta=0$, $z=0$. Wave crests appear to move outwards in the laboratory frame.}
    \label{fig:phase_speed_radial}
\end{figure}

The fact that the oscillation frequency in the comoving frame, $\omegat = \omega - m\Omega(r)$, of the SARIs is appreciable away from the corotation radius (Fig.~\ref{fig:spiral}~(b)) implies that these modes will have non-zero phase speeds. Let us first write the Fourier coefficients in polar notation as
\begin{equation} \label{eq:fourier_polar}
    \hat{f}(r) = p(r)\exp\left(i\int_{r_1}^r q(r') dr' \right),
\end{equation}
where $p,q$ are real functions, and then define the phase for any wave or instability as
\begin{equation} \label{eq:phase}
    \Phi(r,\theta,z,t) = \int q(r)dr + m\theta + kz - \omega_r t.
\end{equation}
The rotational phase speed can then simply be calculated by $v_{\text{ph},\theta} = r\omega_r / m \approx r_* \Omega(r_*)$, which is indeed independent of the sign of $m$. For vertical phase speeds, there is a distinction between $\pm m$ since $v_\text{ph,z} = \omega_r / k = m\Omega(r_*) / k$, so that $m>0$ modes appear upwards-propagating and $m<0$ modes downward propagating for positive $k$. Finally, the radial phase speed can only be defined locally because the radial `wavenumber' $q$ varies over the domain:
\begin{equation} \label{eq:radial_phase}
    v_{ph,r}(r) = -\frac{\partial \Phi}{\partial t} / \frac{\partial \Phi}{\partial r} = \omega_r / q(r) \approx m\Omega(r_*) / q(r_*).
\end{equation}
The radial phase speeds for the SARI from earlier can be read off from Fig.~\ref{fig:phase_speed_radial}, which shows a time-distance diagram of the $B_r$ perturbations along a fixed radial cut as the wave moves through. The trailing nature of the spiral is reflected in the apparent outward-moving phases. Since $\pm m$ modes have similar leading spiral shapes as mentioned in Sec.~\ref{sec:SARI_spirals}, and hence show little differences in their radial phase speeds, they must have similar $q$ but of opposite sign. Around corotation, the radial wavenumber is quite constant, but it increases towards the resonances (not shown), which is also predicted by Eq.~(83) of GK22. This becomes especially clear for modes near marginal stability ($\nu \rightarrow 0$), which oscillate heavily when approaching the singular solutions (GK22), similar to what happens near the corotation singularity in hydrodynamical disks \citep{latter_corotation}. This highlights why a WKB approach cannot properly describe the entire eigenfunction, especially for more localised SARIs near marginal stability, since the radial wavelength at corotation is much larger than that near the resonances, but the mode structure heavily depends on the behaviour near the resonances. 
% For the MRI, the radial wavenumber also varies over the domain \citep{blokland05}, but a WKB approach is valid because they look at the most unstable mode. This will be harder for less unstable modes.

The concept of group velocity is less straightforward to generalise from galactic disks. \citet{toomre69} showed in a local model that wave packages of superposed modes propagate radially and with increasing radial wavenumber as they move away from corotation when assuming a weakly time-dependent wavenumber $q$. The 2D quasi-continua of SARIs contain modes that each already have an inherent wave package-like structure, and that have increasing radial wavenumbers towards marginal stability. Any true transport of angular momentum over the disk by this instability will then also be short-lived, because it will quickly evolve into the non-linear stage. Future work might shed light on this instability, possibly helped by an initial-value approach as in \citet{toomre69}.

With the machinery of the spectral equation \eqref{eq:spectral_equation} very much resembling that of \citet{lynden-bell_ostriker} (the operators in the spectral equation are real and self-adjoint), their `anti-spiral' theorem is readily generalised to ideal MHD: no purely oscillating (i.e. with $\nu=0$) non-axisymmetric mode has a spiral structure. Indeed, any non-degenerate stable, purely oscillating wave will have real solutions for $\boldxi$ and $\omega$. This directly implies that $d\Phi/dr = 0$, so the phase is constant and the mode has no spiral structure. What is different for instabilities like the SARI is that their solution to the spectral equation is necessarily complex, i.e., both the eigenvalue and eigenvector take truly complex values. This causes the spiral shape of the perturbations. The structure of the eigenvalue problem \eqref{eq:spectral_equation} is such that if the pair $(\omega,\boldxi)$ is a solution to the spectral ODE, then so is its complex conjugate counterpart. Hence, any overstability (i.e. growing $\nu\neq 0$ while moving $\omega_r\neq 0$) and its corresponding damped and moving version are closely related. Indeed, reverting to the polar notation \eqref{eq:fourier_polar}, this implies that $q_\text{instab} = -q_\text{stab}$, and hence that they form trailing/leading spiral pairs, as discussed by \citet{lynden-bell_ostriker} for hydrodynamic perturbations. For damped SARI variants, the handedness of Figs.~\ref{fig:spiral} and \ref{fig:phase_speed_radial} would be reversed. The damped mode is hence the time-reversed solution of the growing mode. This time-symmetry aspect was also mentioned for the general equation governing perturbations in ideal MHD, as stated in Appendix C of \cite{keppens2016}.

\section{Instability mechanism} \label{sec:field_topology}
\begin{figure}
    \centering
    \includegraphics[width=0.65\columnwidth]{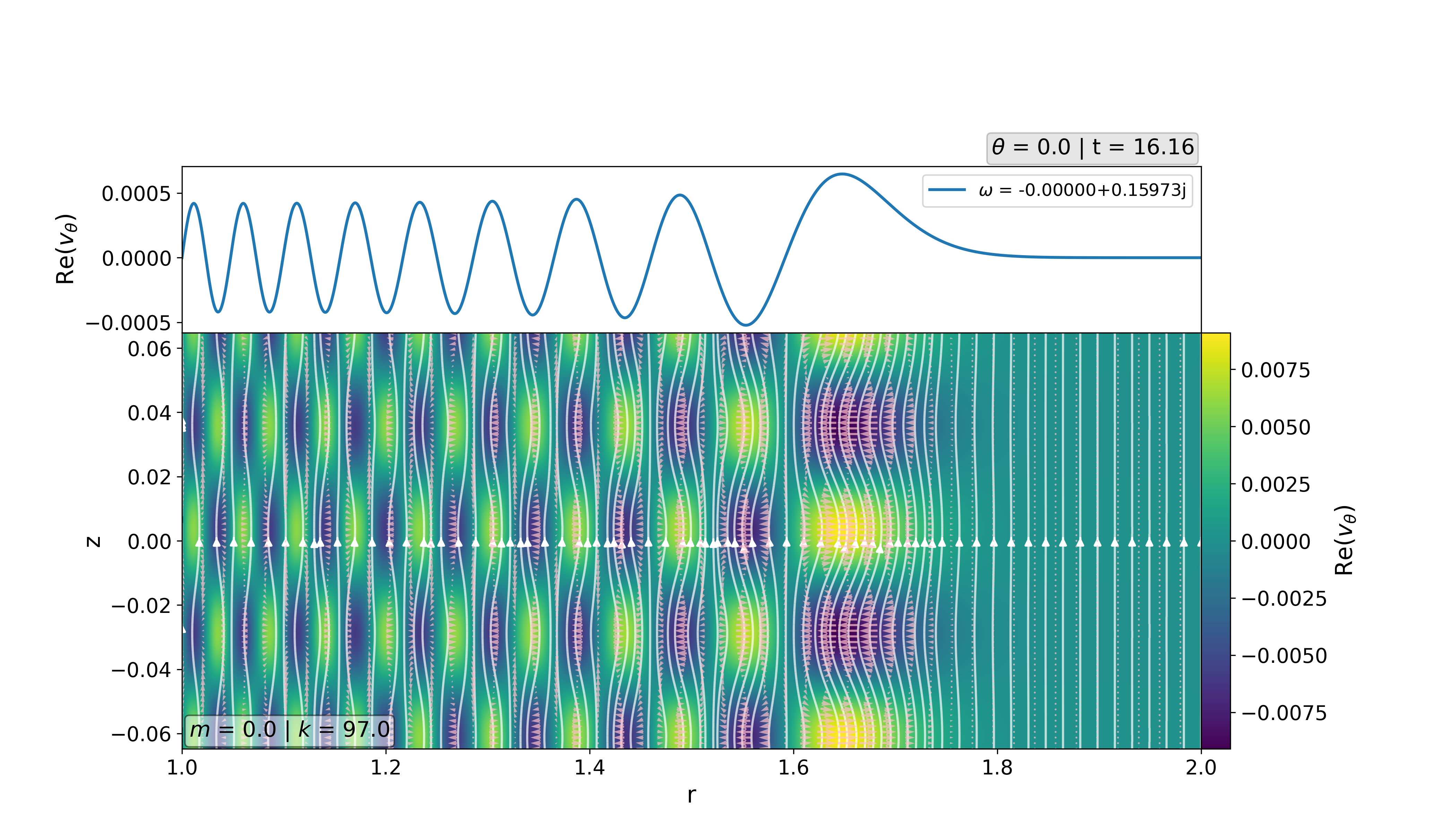}
    \caption{One rather global MRI mode ($m=0$, $k=97$, $\mu_1=0$) shown in a vertical cut along the $r-z$ plane. The projected magnetic field lines are featured on top of the $v_\theta$ perturbation at the end of the linear phase. Quivers denote the projected velocities. An animation is available in the online version of this article.}
    \label{fig:MRI_mechanism}
\end{figure}

The basic MRI operating mechanism is well-understood and nicely illustrated in the lecture notes by \citet{armitage_lecture_notes}. We will first describe how this standard cartoon also arises in a global MRI description, and then generalise it to different field orientations and to the SARI. Figure~\ref{fig:MRI_mechanism} shows a vertical $(r,z)$ cut through the disk, with the projected magnetic field lines superimposed on the $v_\theta(r,z)$ perturbation evaluated at $\theta=0$ of one global MRI mode. Note that $v_\theta$ acts as a proxy for the angular momentum perturbation, since $L(\br,t) = r (v_{\theta0}(\br,t)+v_\theta(\br,t)) = r (r\Omega(\br) + v_{\theta}(\br,t))$. The animated view available online shows how the MRI mode develops in-place without any phase speeds, and deforms the originally purely vertical field and azimuthal flow pattern. Clearly, this more global mode divides the disk in several rings that each produce the same intuitive picture: along a vertical field line, fluid parcels are moved radially inwards in the direction of the flow (positive $\theta$ direction), and radially outwards against the direction of the flow (negative $\theta$ direction). Angular momentum is transferred between these elements by magnetic tension, which is directed opposite to the displacement. As a result, fluid elements that are radially moved inwards are further slowed down with respect to their Keplerian orbital speeds, and the outwards-moved fluid elements are accelerated. Hence, the fact that the fluid and field lines are displaced radially inwards and into the plane (positive $\theta$, ahead of its Keplerian orbit) wherever the (Eulerian) perturbation $v_\theta < 0$, is not contradictory: the fluid then has insufficient angular momentum to stay on that orbit, forcing it to move radially inwards, but also forwards because of the associated Coriolis force. For the radially-outwards-displaced fluid elements, the opposite happens. 

The largest displacement occurs further down the disk, where $v_r$ is maximal. This part of the mode was identified with the local channel flow solution by \citet{Latter2015}, while more inwards perturbations were described as a `radially-varying' mode. The global picture presented here indeed matches that description (and in particular their Fig.~(1)), where the radially-varying part of the mode has nodes with counter-streaming velocities at either sides, and purely vertical motion at the nodes themselves. In this regard, these radially-varying parts are different from the channel modes parts (which have zero radial wavenumber and no vertical flow), but \citet{Latter2015} showed that these radially-varying regions can be mapped to a channel mode at a different radius, with the same growth rate after rescaling. Hence, very similar dynamics occur at various locations over the disk, where the inner parts of the `radially-varying' mode can in fact be seen as small-scale channels that are impeded in their growth by the nodes of the mode, where the displacements are in the vertical direction. Note that, in an MHD spectroscopy approach, the number of discrete MRI modes at a particular vertical wavenumber $k$ is finite and the modes are connected to at least the inner boundary \citep{GK22}. They are in fact part of an infinite sequence of modes that continues into oscillating modes on the real frequency axis, clustering to the Alfvén/slow continua. The finitely many MRI are wall-influenced and become more global (with increasing radial wavenumber) as the number of nodes increases as we descend along the countable sequence of modes. Especially for the MRIs with smaller growth rates, the eigenfunctions can be truly global and span the entire radial extent of the disk, connecting to the outer boundary. In such a case, the `true' channel part of the mode gets lost and only `radially-varying' parts remain. Even though such a global MRI would locally behave like a channel mode, it is unclear whether this behaviour continues in the non-linear regime. These local channels might interact and quickly break any channel structure because of the radially opposite displacements. In this way, we agree with the conclusion of \citet{Latter2015} that the abundance of channel modes in shearing box simulations might be artificial. 

\subsection{Polarization of $\boldB$ and $\boldv$ for MRI versus SARI} \label{sec:efs_numerical}

\begin{figure}
    \centering
    \begin{subfigure}{0.49\columnwidth}
        \includegraphics[width=0.95\linewidth]{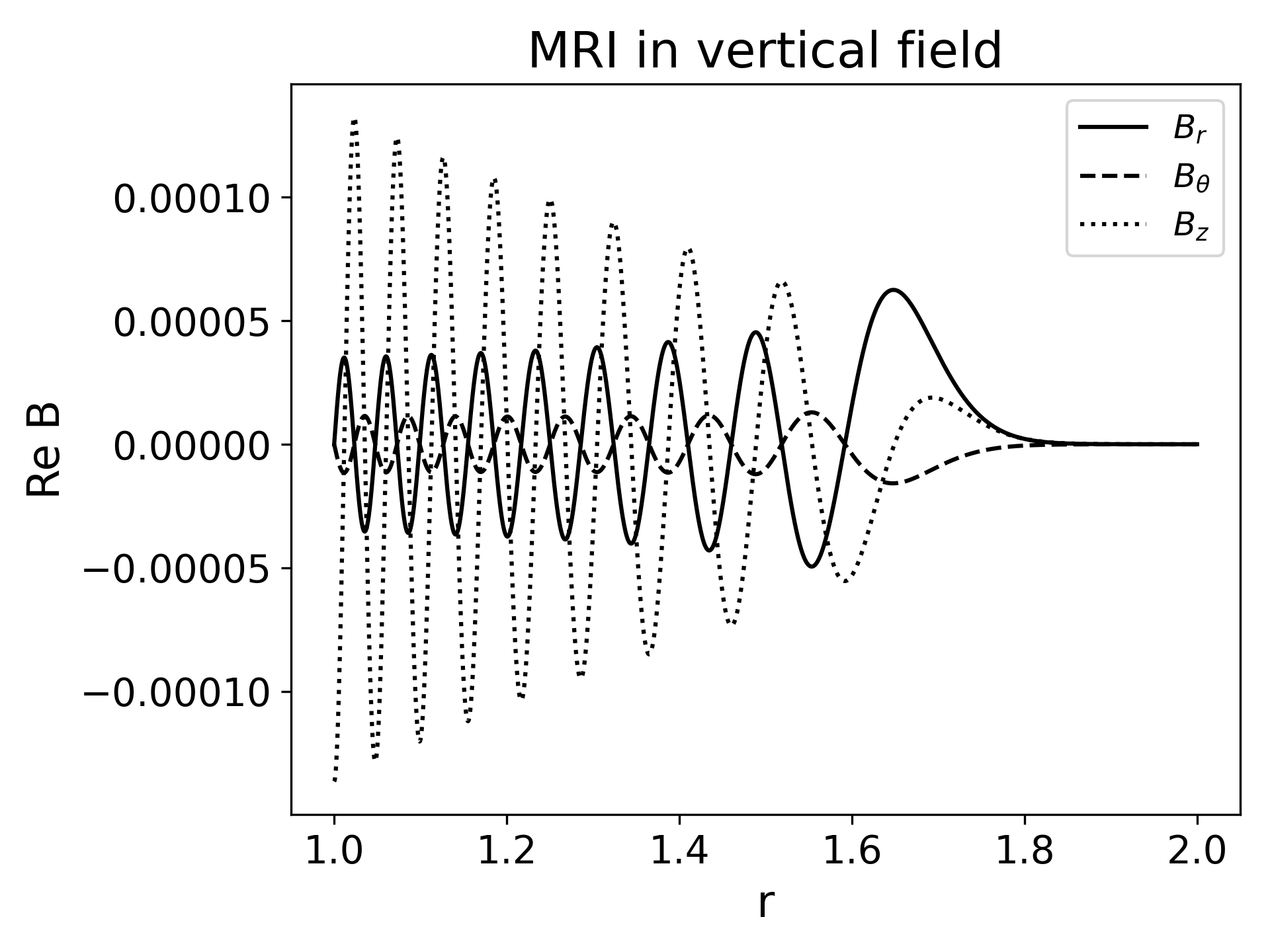}
    \end{subfigure}
    \begin{subfigure}{0.49\columnwidth}
        \includegraphics[width=0.95\linewidth]{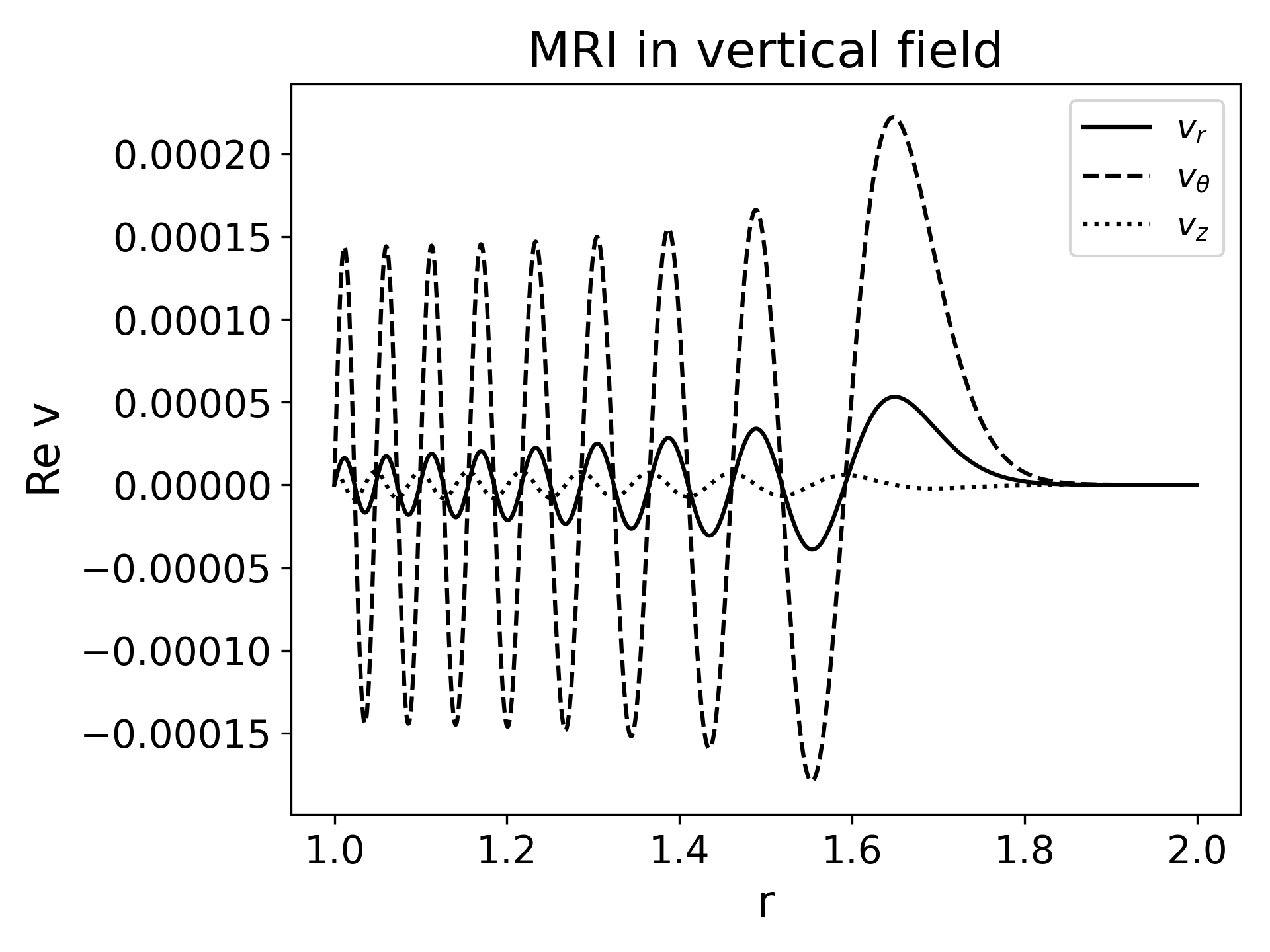}
    \end{subfigure}
    \caption{(a) Relative amplitudes and orientations of the perturbed magnetic field components in cylindrical coordinates for the MRI in a vertical background field ($m=0$, $k=97$, $\mu_1=0$, $\omega = 0.1597i$). $B_r$ and $B_\theta$ are in anti-phase, and $90^\circ$ out-of-phase with $B_z$. The field-aligned component dominates in amplitude, indicating a predominant slow nature of the mode. (b) Similar as in (a) but for the velocity field.}
    \label{fig:field_efs_MRI}
\end{figure}

More direct insight into the nature of the instability and the polarization of the field can be obtained from the eigenfunctions for $\boldB$ and $\boldv$ themselves. In this Section, the quantities $B_{r,\theta,z}$ and $v_{r,\theta,z}$ denote the perturbed components. Fig.~\ref{fig:field_efs_MRI}~(a) and (b) show the real parts of these eigenfunctions for the same MRI mode as above in Fig.~\ref{fig:MRI_mechanism}. Taking the real parts amounts to considering the perturbations at $t=0$, $\theta=0$, $z=0$. By focusing in particular on the zeroes and extrema of the eigenfunctions, the obtained results are applicable to every location $\theta,z$.

First note how the basic mechanism of Fig.~\ref{fig:MRI_mechanism} also appears from Fig.~\ref{fig:field_efs_MRI}: the perturbations for $B_r$ and $B_\theta$ are in anti-phase, whereas $v_r$ and $v_\theta$ are in phase. This matches with the idea of oppositely-moving fluid elements. It is approximately true (see Sec.~\ref{sec:efs_analytical}) that both $B_z$ and $v_z$ are zero whenever the radial and azimuthal components are maximal, and vice-versa. We checked this for various field orientations, and the zeroes of $B_r'$ and $B_z$ usually match up to $10^{-3}$. Remarkably, the zeroes of the components of $\boldB$ and $\boldv$ match closely. This behaviour essentially appears because discrete MRI eigenfunctions are almost real, up to rotation. That means that they have true zeroes, in the sense that the modulus of the eigenfunctions has zeroes. Note how the $B_z$ eigenfunction is maximal near the inner edge of the disk and decreasing, whereas $B_r$ is maximal in the interior of the disk, generalising the result of \citet{Latter2015} to equilibria with radially varying magnetic field profiles.

\begin{figure}
    \begin{subfigure}{0.49\columnwidth}
        \includegraphics[width=0.95\linewidth]{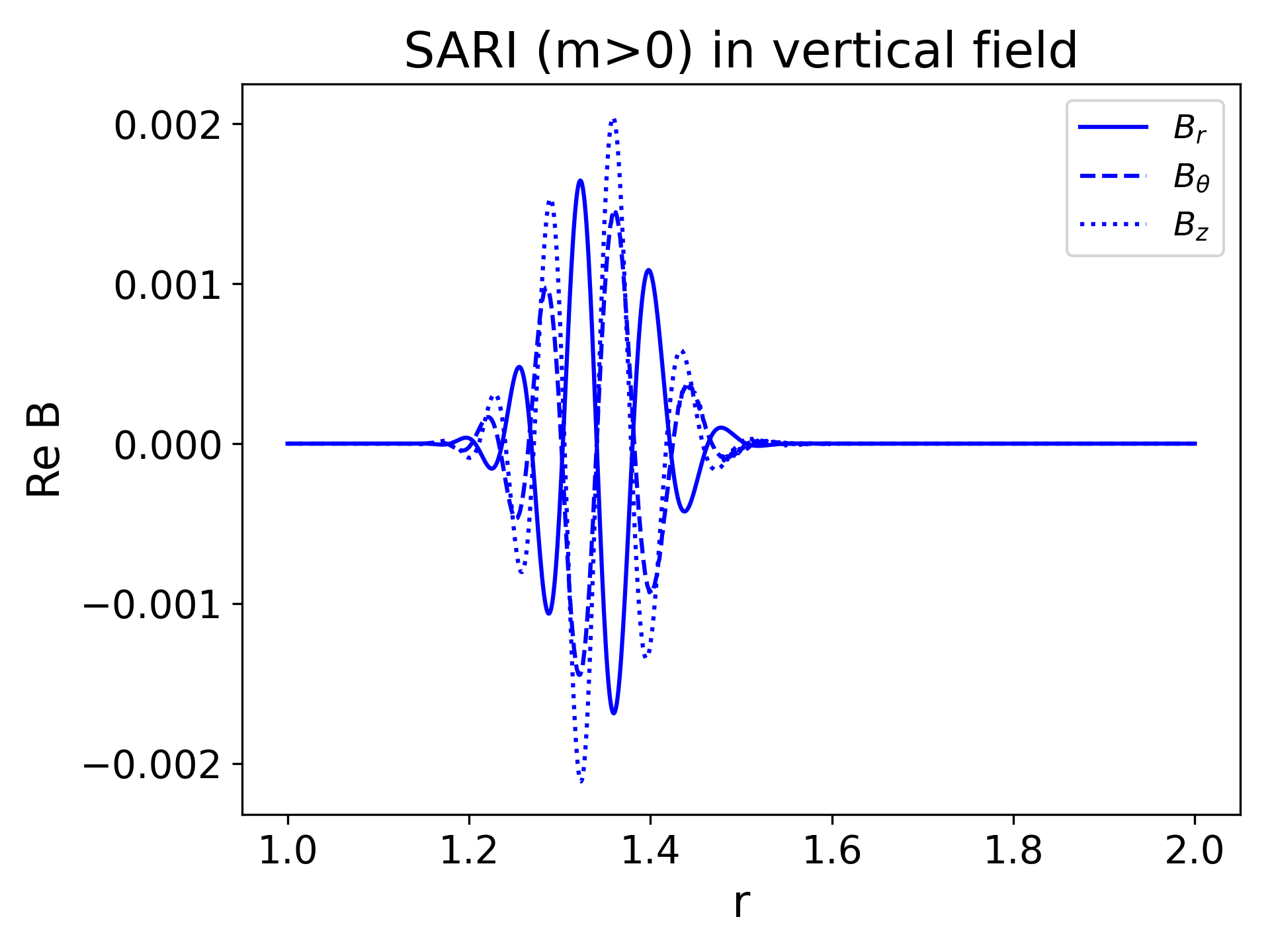}
        \label{fig:field_efs_SARI_co}
    \end{subfigure}
    \begin{subfigure}{0.49\columnwidth}
        \includegraphics[width=0.95\linewidth]{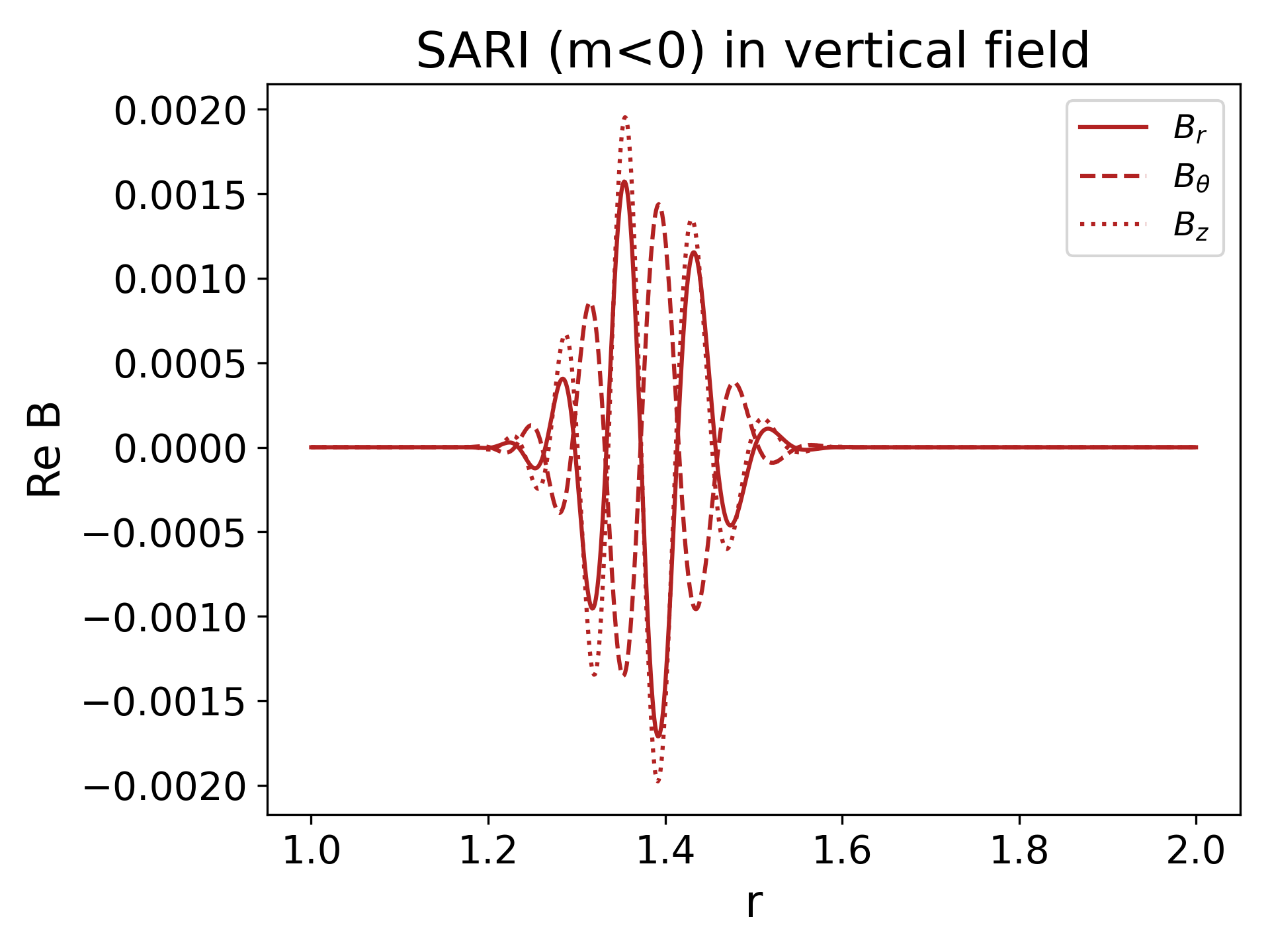}
        \label{fig:field_efs_SARI_counter}
    \end{subfigure}
    \caption{Relative amplitudes and orientations of the perturbed magnetic field components for the SARIs ($m=\pm2$, $k=65$, $\omega=1.28533+0.24458i$ and $-1.23406+0.24240i$) in a vertical background field ($\mu_1=0$). $B_r$ and $B_\theta$ are in anti-phase, and $B_z$ is in phase or in anti-phase with one of them. Again, $B_z$ dominates because of the slow nature of the modes.}
    \label{fig:field_efs_SARI}
\end{figure}

The polarization of $\boldB$ for $\pm m$ SARIs in a vertical field is shown in Fig.~\ref{fig:field_efs_SARI}. Note how the eigenfunctions are localised around corotation. Again, $B_{r,\theta}$ are in anti-phase and $v_{r,\theta}$ in phase. The situation is now quite different compared to the MRI, since the $B_z$ perturbation is either in phase ($m<0$) or in anti-phase ($m>0$) with $B_\theta$ (the sign combination matches that of the wavenumbers $m,k$). The eigenfunctions are now really complex functions and have no zeroes in the domain, which becomes apparent by plotting the modulus of the eigenfunctions. This implies that the zeroes of the real and imaginary parts always interlace. 

An overview of the polarization of MRI and SARI eigenfunctions is given in Table~\ref{tab:polarizations}. Remarkably, the results are the same for the four background field orientations under consideration. The quantifications were made around the corotation radius for the SARIs, and near the `channel' part of the MRI. The imaginary parts show a very similar behaviour, but with a flipped increasing/decreasing behaviour around the zeroes. We explain these polarization properties analytically in Appendix~\ref{sec:efs_analytical}.

\begin{table}
	\centering
	\begin{tabular}{lcccccc} % 
		\hline
		 & $B_r$ & $B_\theta$ & $B_z$ & $v_r$ & $v_\theta$ & $v_z$ \\
		\hline
		MRI & $+$ & $-$ & $0\downarrow$ & $+$ & $+$ & $0\downarrow$\\
		MRI damped & $+$ & $+$ & $0\uparrow$ & $+$ & $+$ & $0\uparrow$\\
		SARI+ & $+$ & $-$ & $-$ & $0\uparrow$ & $0\uparrow$ & $0\downarrow$\\
        SARI+ damped & $+$ & $+$ & $+$ & $0\uparrow$ & $0\downarrow$ & $0\uparrow$\\
        SARI- & $+$ & $-$ & $+$ & $0\downarrow$ & $0\downarrow$ & $0\downarrow$\\
        SARI- damped & $+$ & $+$ & $-$ & $0\downarrow$ & $0\uparrow$ & $0\uparrow$\\
		\hline
	\end{tabular}
	\caption{Polarization of MRI and SARI eigenfunctions in different field orientations. Only the real part is considered here. Whenever the real part is zero, the imaginary part will be zero for the MRI (up to rotation), and near an extremum for SARIs. We compare the location where the real radial component is maximal ($+$) to the other components, which can be minimal ($-$) or increasing or decreasing through zero $(0\uparrow, 0\downarrow)$.}
	\label{tab:polarizations}
\end{table}

For instabilities in any background field, the polarization is indeed such that $B_r$ and $B_\theta$ are in anti-phase. This is in agreement with the MRI cartoon and basic understanding in terms of angular momentum, since then an inward-moving fluid parcel will lose angular momentum and move faster with respect to its original radius, and vice-versa for an outward-moving fluid parcel. This is opposite for the damped version of MRI/SARI: there, the $B_r$ and $B_\theta$ components are in phase, so the fluid parcel that was moved radially outwards is displaced ahead of the background flow, which is a perturbation that will be stabilised by the same angular momentum argument (here, the magnetic tension force results in a transfer of angular momentum between fluid parcels that restores them to their equilibrium orbit. The associated Coriolis force acts to undo the displacement in the $\theta$ direction). 

In terms of magnetic field perturbations, the difference between MRI and $\pm m$ SARIs persists for all background fields. For the MRI, the $B_z$ perturbation is out-of-phase with the other components, which means that it vanishes where $B_{r,\theta}$ have their extrema. This means that the total field only has an upwards component at the locations where it is maximally displaced, as we mentioned with respect to Fig.~\ref{fig:MRI_mechanism}. For the $m>0$ SARI, $B_z$ is in anti-phase with $B_{r,\theta}$, while for the $m<0$ SARI, $B_z$ is in phase with $B_{r,\theta}$. Hence, for both these modes the field line deformations will be modified with respect to the MRI: in a vertical field, the additional perturbed $B_z$ component pairs the inwards and outwards-displaced fluid elements together, either with the outwards part on top ($m>0$) or with the inwards part on top ($m<0$) (see Fig.~\ref{fig:SARI_mechanism} for an example). For a helical field, the behaviour will be similar, but the up- or downward displacement of the field lines becomes more prevalent. For a purely azimuthal field, the opposite `handedness' of the perturbations will become obvious in the vertical displacements of the field lines (see also Fig.~\ref{fig:field_lines}~(d), which shows perturbations of a nearly azimuthal field). The sign of $k$, like $m$, also determines whether $B_z$ will be in phase or in anti-phase with $B_r$. As a general rule, whenever $k/m$ is positive, $B_z$ will be in phase with $B_\theta$. The displacement of the field lines is hence partly determined by the direction of the wavevector.

The velocity field perturbations show similar behaviour as the magnetic field perturbations, but $v_r$ and $v_\theta$ are always in phase. The same angular momentum argument as for the instability of the MRI can be applied to the SARI, but now, magnetic tension has an additional vertical component. The phase of $v_z$ is the same as that of $B_z$, which in turn is in (anti-)phase with $v_\theta$ depending on the sign of $m$. Quantifying the angle $\psi$ between $\boldB$ and $\boldv$, we find that for the more global MRI it varies over the domain, but it is approximately $90^\circ$ near the channel part of the mode. \citet{latter09} state that $\psi$ only deviates significantly from this value as a result of resistivity (e.g. in protostellar disks), but this clearly only applies to the part of the disk where the perturbation is maximal. For SARIs, we find an angle of almost exactly $90^\circ$ near corotation: indeed, the components of $\boldB$ and $\boldv$ are out of phase in Table~\ref{tab:polarizations}. This observation also points to the combined slow-Alfvénic nature of the modes under nearly incompressible conditions.

There is a small variation in terms of angle that the perturbed field itself makes with the radial line, but it is usually around $\pi/4$, in correspondence with what \citet{latter09} identify as the fastest-growing channel. For the more global MRI modes, the value of $\tan(B_r/B_\theta) \approx -1$ is decreasing over the domain. For discrete MRI modes with lower growth rates (i.e. the more global ones), the angles tend to increase. Indeed, approaching marginal stability means that the mode gets closer to the polarization of an ordinary oscillating slow/Alfvén mode. That means that $B_r$ disappears completely at $\nu\rightarrow0$ in favour of $B_{\theta,z}$, resulting in an angle of $90^\circ$ with the radial line. In this situation, the field is just bent in the direction of the flow, producing a traveling wave that is driven by field tension, not an instability thriving on the balance between centrifugal and tension forces. 

Comparing the relative amplitudes of the perturbations for the different background fields, it turns out that the field-aligned component of $\boldB$ always dominates. This aligns with the usual interpretation the MRI as unstable slow modes: it is possible to write a perturbation in terms of a coordinate system aligned with the background field, $\boldxi = (\xi_r, \xi_\parallel, \xi_\perp)$. The basic result of MHD waves in a homogeneous medium that gives the polarization of the Alfvén mode in purely $\xi_\perp$, the slow mode purely $\xi_\parallel$, and the fast mode purely $\xi_r$, is transferred in some sense to inhomogeneous MHD \citep{GKP19}. Although the modes are hybrid (they show mixed properties since both MRI and SARI have all three components), the $\xi_\parallel$ component dominates, which points to a predominantly slow mode. In fact, when going towards marginal stability ($\nu \rightarrow 0)$ for both MRI and SARI, the polarization becomes more extreme in terms of $\xi_\parallel$ as the modes get closer to the polarization of pure slow modes with additional Alfvénic properties. As was mentioned above, the polarization properties associated with the MRI and SARI modes are analytically reproduced in Appendix~\ref{sec:efs_analytical}.

\subsection{Visualising 3D magnetic field line deformations}

\begin{figure}
    \centering
    \includegraphics[width=0.65\columnwidth]{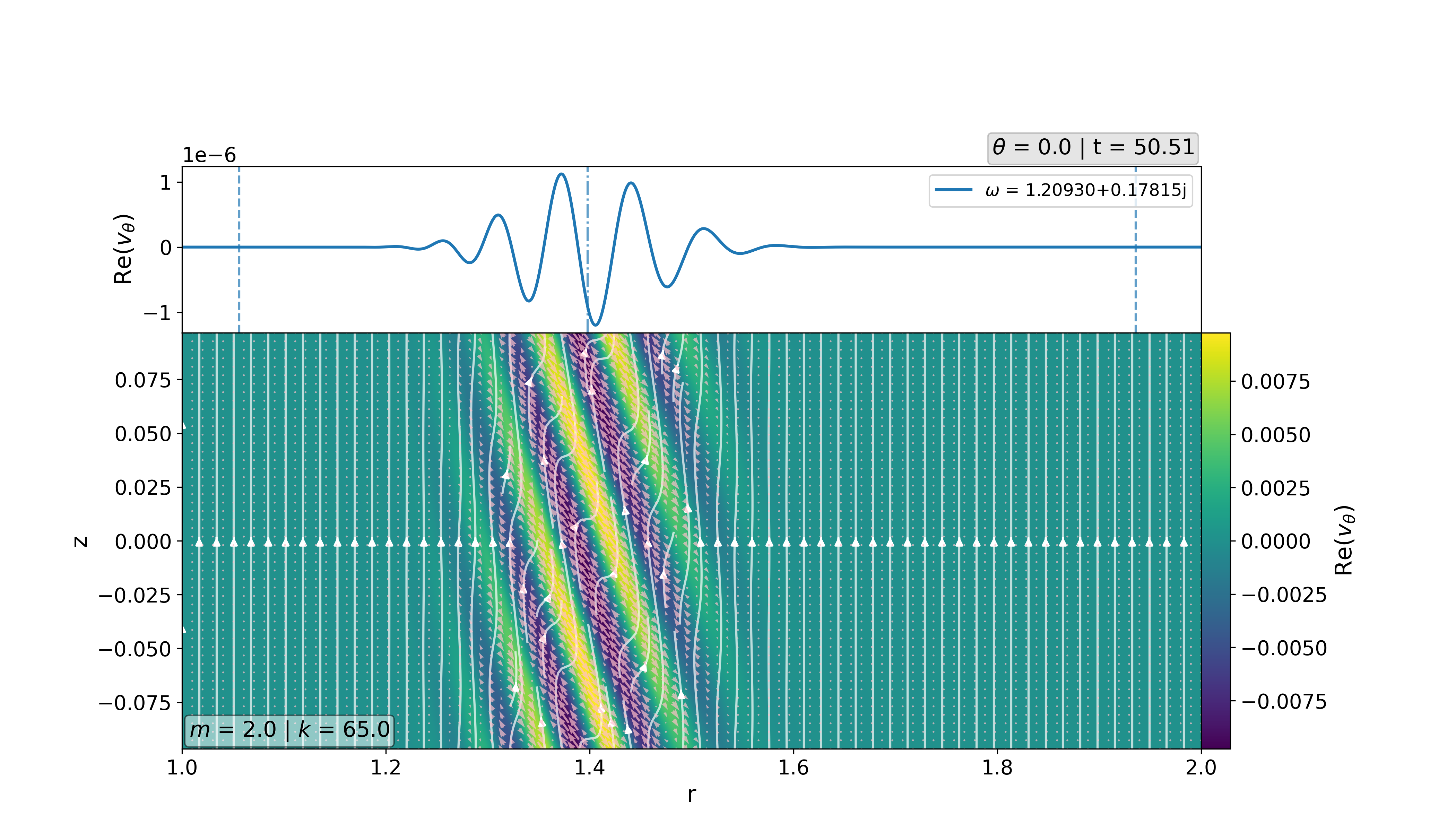}
    \caption{One SARI mode ($m=2$, $k=65$, $\omega = 1.20930+0.17815i$) in a vertical field ($\mu_1=0$), shown in a vertical cut along the $r-z$ plane. The projected magnetic field lines are shown on top of the $v_\theta$ perturbation at the end of the linear phase. Quivers denote the projected velocities. An animation is available in the online version of this article.}
    \label{fig:SARI_mechanism}
\end{figure}

We now look specifically at how the magnetic field lines are perturbed for the MRI/SARIs in different background fields. The field lines for a vertical background field are appropriately shown in a 2D visualisation like Fig.~\ref{fig:MRI_mechanism} for the MRI or Fig.~\ref{fig:SARI_mechanism}, which shows a SARI mode. For MRI, the field lines are a 2D projection because of the axisymmetry. For the SARI, the variation in the $\theta$ direction is sinusoidal, and hence it would a priori be false that the 2D field lines are projections of 3D curves. However, because of the low $m=\pm 2$ in this visualisation, non-axisymmetry is small on the scale of the perturbed field lines and they approximate 2D projections. Figure~\ref{fig:SARI_mechanism} shows the perturbed field lines for the $m=2$ SARI whose eigenfunctions were shown above. Note how, compared to the MRI picture in Fig.~\ref{fig:MRI_mechanism}, the field is now indeed also vertically displaced. Also note how the contours of constant phase point along the wavevector and in the direction of the field displacements. The instability mechanism is the same as that for the MRI: the displacements of the field lines are associated with the maximal radial and azimuthal velocities, and hence exchange angular momentum between fluid elements. In contrast to the MRI, the angular momentum exchange is now not purely radially, but also vertically. The field lines for $m<0$ SARIs are not shown, but they are mirrored versions of those on Fig.~\ref{fig:SARI_mechanism}, as are the lines of constant phase, which in that case point diagonally upwards. We also checked that for negative $k$, the perturbations are mirrored about $z=0$, so that indeed the identification with the direction of the perturbation and the wavevector can be made: the perturbed velocity field is parallel to the wavevector $\vec{k} := (q, m/r, k)$. This contrasts with the MRI picture, where the magnetic field is displaced only locally radially inwards and outwards, without overall radial drift.

\begin{figure}
    \centering
    \includegraphics[width=0.7\linewidth]{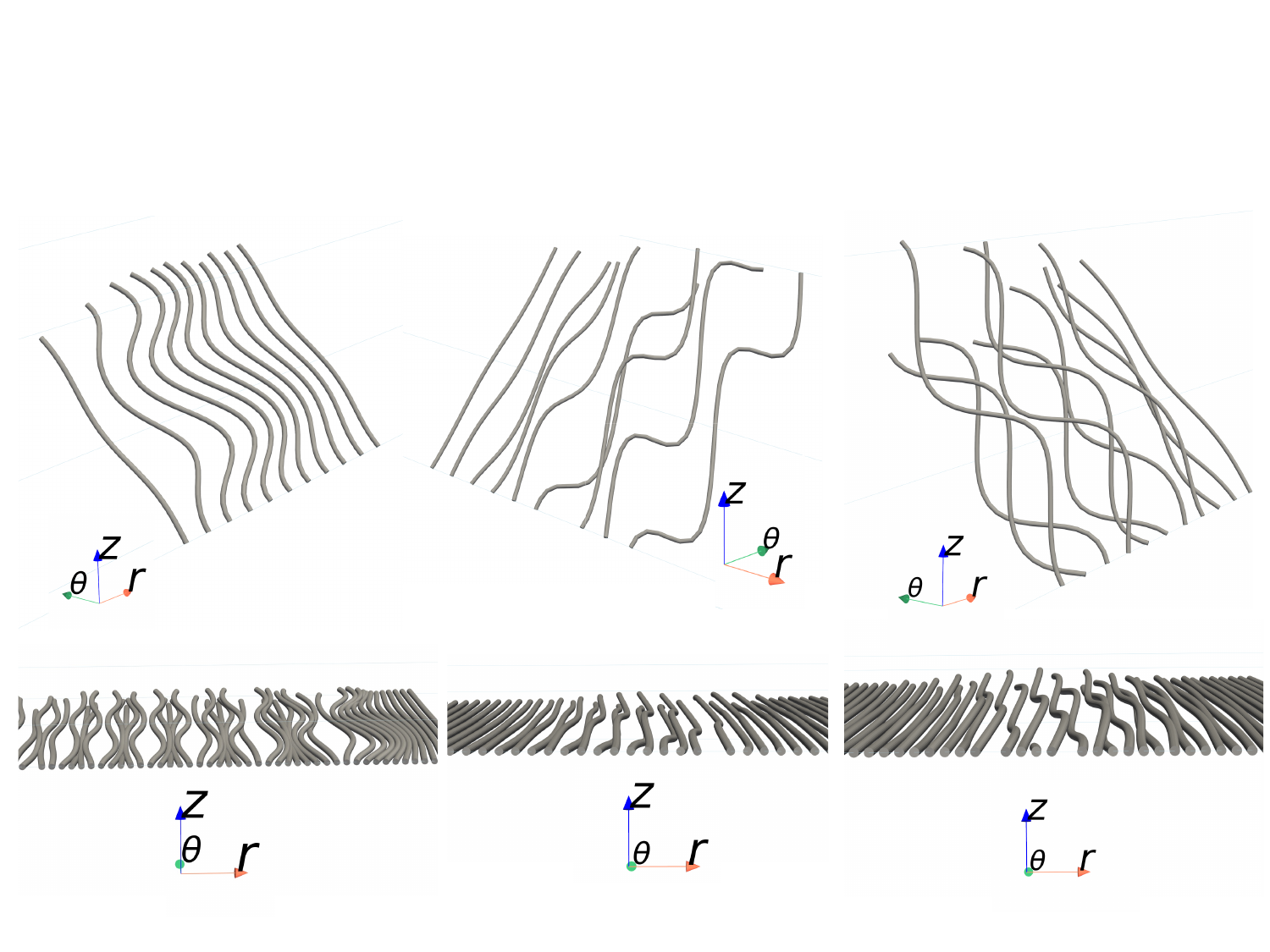}
    \caption{Field lines for the MRI and $m=2$ and $m=-2$ SARIs, respectively, in a helical field ($\mu_1=1$, top row, with wavenumbers $k=137$, $90$, and $93$, and eigenvalues $-0.00074+0.25330i$, $1.17589+0.22422i$, $-1.16113+0.2236i$, resp.) and nearly azimuthal field ($\mu_1=10$, bottom row, with wavenumbers $k=972$, $635$, and $662$, and eigenvalues $-0.00150+0.46723i$, $1.16981+0.20456i$, $-1.15691+0.17374i$, resp.). Notice how the displacement is mostly radially for the MRI, but has a clear vertical component for the SARI.}
    \label{fig:field_lines}
\end{figure}

For field orientations other than vertical, it is expedient to go to 3D visualisations. Figure~\ref{fig:field_lines} shows field lines for the MRI/SARIs in a helical field and in a nearly azimuthal field. Two main observations can be made: (1) comparing the MRI to the SARI, the maximal displacements are again radially versus diagonally, even in dominant azimuthal fields; (2) the `handedness' of the SARI field is clearly visible.

\begin{figure}
    \centering
    \includegraphics[width=0.65\columnwidth]{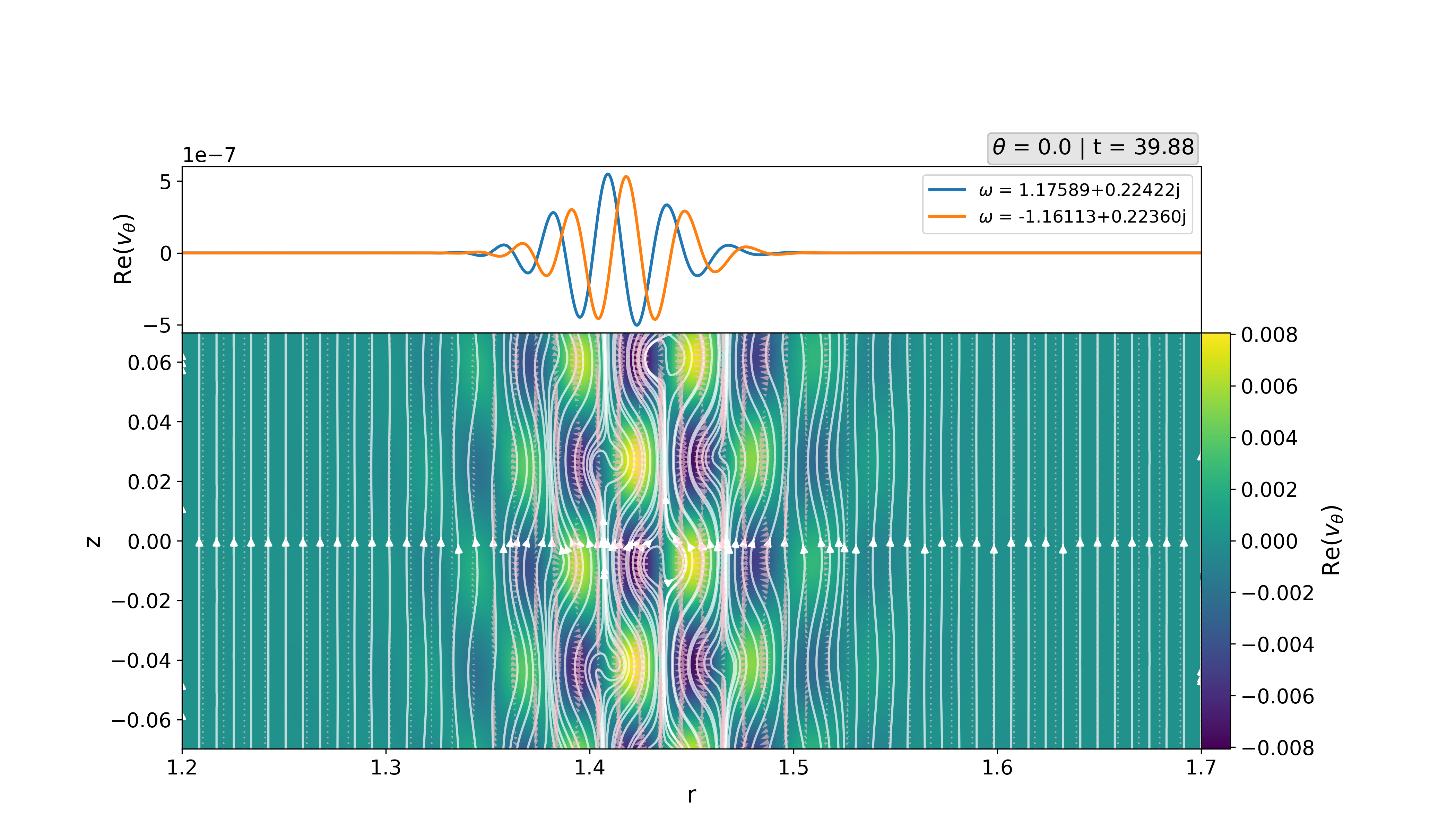}
    \caption{Two slightly misaligned SARI modes with opposite $m$ ($m=2$, $k=90$ and $m=-2$, $k=93$) and similar growth rates ($\omega=1.17589+0.22422i$ and $\omega=-1.16113+0.22360i$) result in the formation of island-like structures in a vertical cut along the $r-z$ plane (helical field $\mu_1=1$). The projected magnetic field lines are shown on top of the $v_\theta$ perturbation at the end of the linear phase. Quivers denote the projected velocities. An animation is available in the online version of this article.}
    \label{fig:SARI_islands}
\end{figure}

The fact that the displaced field extrema for the SARI have an upwards component along the wavevectors opens the door for interesting dynamics. To illustrate that, Fig.~\ref{fig:SARI_islands} shows two superposed SARI modes with opposite $m$ -- and hence an opposite direction for the wavevector and field displacements -- that are slightly misaligned. The result of this misalignment is that, instead of canceling out everywhere, field topologies with abruptly reversing field direction occur locally in the `nodes' of the new mode towards the end of the linear phase. These structures are reminiscent of the onset of reconnection to produce magnetic islands and are impossible to obtain with axisymmetric MRI modes alone. Without fully non-linear, resistive simulations, it is impossible to know whether such structures will reconnect, and what their contribution might be to small-scale plasmoid formation and heating of the disk \citep{ripperda2020}. The animated version of Fig.~\ref{fig:SARI_islands}, available in the online version of the paper, clearly shows how the phase shifts of the modes make that the nodes appear outwards-moving. The background field in Fig.~\ref{fig:SARI_islands} is helical ($\mu_1=1$), but we checked in a 3D visualisation that the 2D projection captures the essential behaviour of the field lines. The islands appear as flux tubes in a 3D visualisation, reminiscent of the initial conditions that are sometimes used in shearing box simulations \citep{hirose2006, simon2012}. It is to be emphasised here that these islands and flux tubes occur by a mere linear superposition of two (of the double infinitely many) quasi-continuum SARIs of opposing $m$ value, and that both such modes are linear (near-)eigenmodes of ideal MHD that would need some minute energy addition channel.

\section{Global disks} \label{sec:many_modes}

Up to Fig.~\ref{fig:SARI_islands} where two SARI modes were superposed, all prior figures only incorporated a single pure MRI or SARI eigenmode to highlight similarities as well as their distinctive characters. With the differences in visual appearance and polarization of global MRI and SARI modes established, we now look at how a global disk, where realistically many modes coexist at once, appears in a synthetic visualisation. To separate the effects of MRI and SARI, we artificially superimpose only modes of the same kind ($m=0$ or $m\neq0$) and visualise these global disks as well as determine the stresses associated to the perturbations. 

\subsection{Disks containing multiple MRI or SARI modes}

As mentioned in Sec.~\ref{sec:equilibrium}, we need to remain consistent with the assumption of vertical localisation and hence we need to take into account restrictions on the vertical wavenumbers $k$ such that the modes fit in the vertical scale height. We compare disks where many axisymmetric MRI modes have been excited versus disks where many non-axisymmetric SARIs have been excited by superposing the Fourier eigenmode solutions for many values of $(m,k)$. Given values for $m \in \{-20, \ldots, 20\}$, these $k$ were randomly selected in a range around the wavenumber of maximal growth $k_\text{max}$, determined by $\omega_A = \sqrt{15/16} \Omega$ \citep{BKG24}. For the SARIs, we only want to select quasi-continuum modes that are confined between their virtual walls and not attached to the true walls, which are artificial in any disk. Hence, at the same time, $k$ has to be chosen such that a quasi-continuum of modes exists, which means that the criterion $1 \ll |k/m| < B$ needs to be satisfied (Eq. 91 of GK22) in order for the Alfvén continua to overlap. Given the three background fields described earlier (vertical, helical and almost azimuthal), we generate six datasets in this fashion, while keeping the other equilibrium quantities constant ($B_0 = 0.01$, $\beta=10$ and $r_2-r_1=1$) in the thin disk regime as before. Each mode receives a randomised complex rotation and a small semi-randomised amplitude, chosen such that the total growth at the end of the linear phase is somewhat uniform in the entire disk (the maximal local growth is given by $\sim 0.75 \Omega(r)$, so modes in the outer disk are necessarily slower-growing than modes in the inner disk). The total perturbation is then normalised such that the initial field perturbation is maximally $10^{-4} B_0$. Because the natural growth rates are decreasing with radius, the duration of the linear phase as defined in Sec.~\ref{sec:defining_linear} is determined by the inner disk. Since it is then impossible to show the disk in a fully non-linear state, we show it at an intermediate time $t_\text{max}/4$ where the full disk is still in the linear phase. Per wavevector $(m,k)$, we select those modes with growth rates $> 0.02$ from the finite number of MRIs. For the vertical field, the number of values for $k$ is $n_k=7$, yielding $n_\omega=111$ different modes. For the helical field, $n_k=5$ and $n_\omega=130$ while for the nearly azimuthal field, $n_k=10$ and $n_\omega=1100$ because the modes have a large radial wavenumber and hence it takes many of them to cover the global disk. For the SARI datasets, we apply the Arnoldi shift-invert algorithm around a certain complex frequency in the quasi-continuum regime for every wavenumber combination. This gives $n_k=19$, $n_\omega=221$ for the vertical field, $n_k=40$ and $n_\omega=420$ for the helical field, and $n_k=218$, $n_\omega=7899$ for the nearly azimuthal field. Again, the final case requires more modes to cover the disk because they are more radially localised. All $m_j/k_j/\omega_j$ combinations along with $c_j$ are stored and reproducible.

\begin{figure}
    \includegraphics[width=\linewidth]{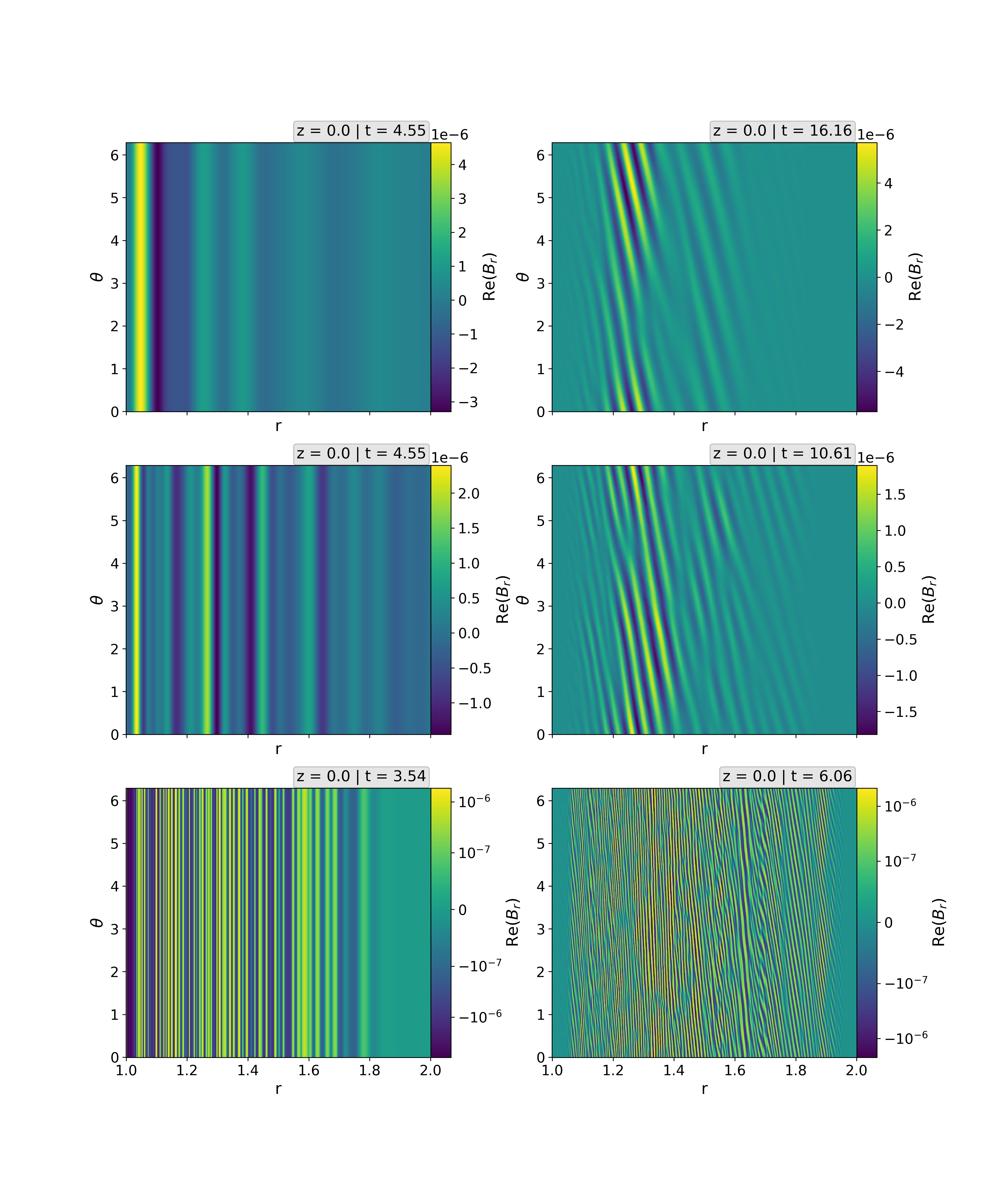}
    \caption{Cylindrical slices taken at $z=0$ of the perturbed radial magnetic field for MRI modes (first column) and SARI modes (second column) at many wavenumbers $m,k$ at three field orientations. Top row: vertical field, middle row: helical field ($\mu_1=1$), bottom row: nearly azimuthal field ($\mu_1=10$). The time is determined by $t_\text{max}/4$. The bottom panels are shown in log colourscale to improve visibility of the fine radial structure.}
    \label{fig:many_modes}
\end{figure}

Figure~\ref{fig:many_modes} features the six datasets in an $r-\theta$ view, with the MRI disks in the first column, and the SARI disks in the second column. The three rows, respectively, correspond to the three field orientations (vertical, helical $B_{\theta1}/B_{z1}=1$, and nearly azimuthal $B_{\theta1}/B_{z1}=10$). The first column clearly shows how the MRI consists of global modes that perturb the entire disk from inner to outer radius. The finite number of MRI modes at a given wavenumber $k$ consist of a collection of discrete modes that are connected to at least the inner boundary, with the most unstable mode localised around the inner boundary and the slower-growing modes mode global over the entire radius. The banded structure is a result of axisymmetry, which is clearly artificial in a realistic disk, and has to be broken through the non-linear evolution of these modes or by external perturbations. The SARI modes, in contrast, are chosen such that they are all detached from the walls (the perturbation close to the walls is in fact exactly $0$), and show fine spiral structure. Unlike the MRI, true non-axisymmetry in this fine structure is present from the very start when these linear modes are excited. When the azimuthal field component increases for both MRI and SARI, the modes become more localised and a finer radial structure is visible: the radial `wavelengths' are smaller. For the SARI, the direct reason is that as $B_\theta$ increases relative to $B_z$, the Alfvén continua overlap more and this results in resonances that are closeby, radially confining the eigenfunctions to smaller scales. The spiral appearance of the modes is hence more tightly wound. The vertical wavenumber $k$, which is large for the more azimuthally oriented fields to be close to maximal growth, plays a similar role. This also explains why the MRI is more localised: from the usual WKB dispersion relation for the MRI (see e.g. \citet{GK22}, Eq.~(25)), it follows that only a range of wavenumbers $q/k$ is allowed for instability. Assuming that $\omega_A\approx \Omega$ so that the MRI growth is maximal and hence $\nu \in [0, 0.75\Omega]$, this gives an allowed range of $q/k\in[0,\sqrt{2}]$. Since larger $k$ is needed to be close to maximal growth for a stronger azimuthal field, this means that $q$ can also become larger, so that the radial wavelengths are smaller. 

\begin{figure}
    \centering
    \includegraphics[width=\linewidth]{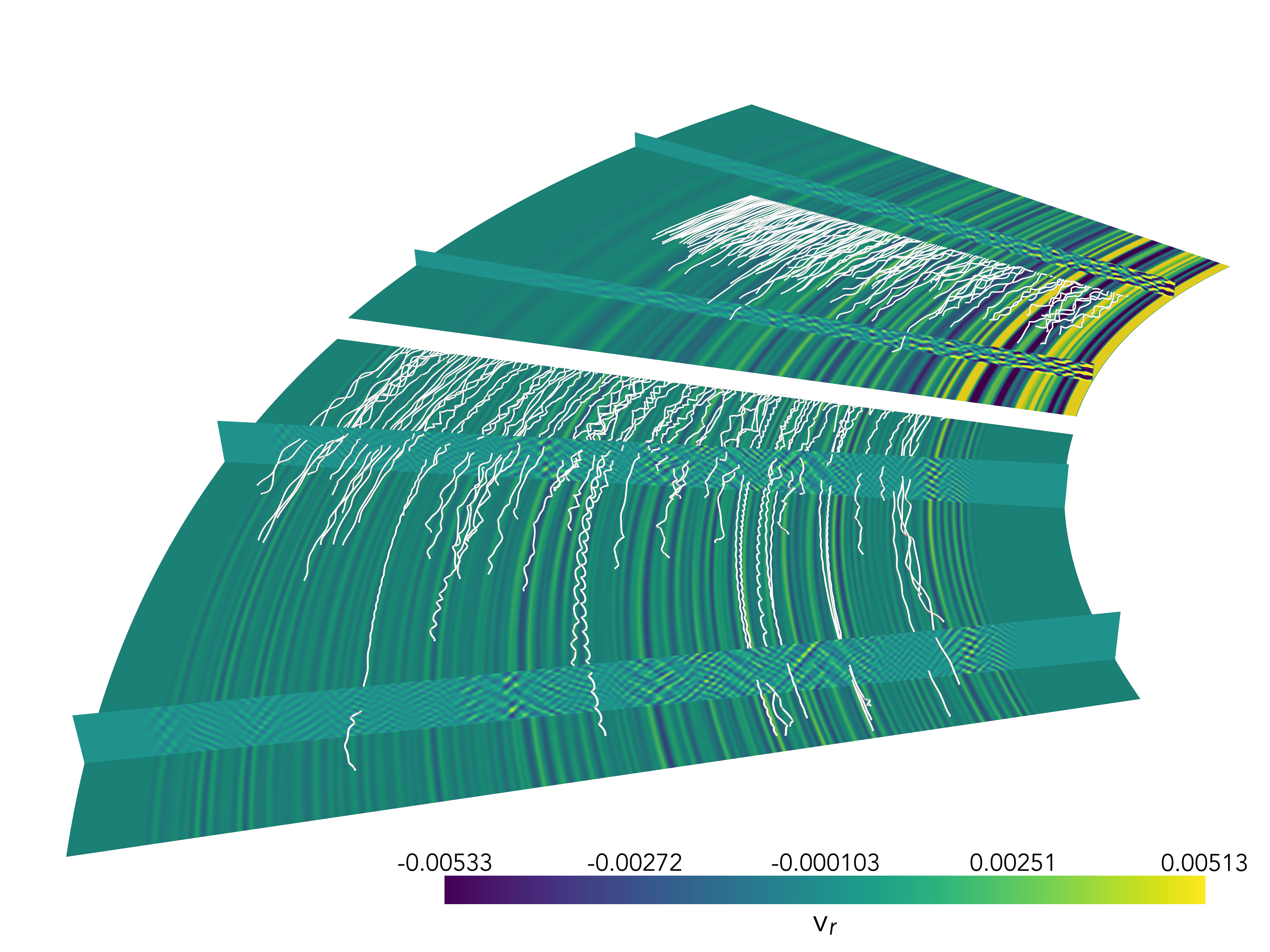}
    \caption{Slices of a 3D MRI (back) and SARI (front) disk with modes at many wavenumbers and $\mu_1=10$ at time $t_\text{max}/8$. Colorbar shows the perturbed $v_r$ component. Field lines are shown in white. Note how the SARI field has a larger vertical displacement compared to the MRI. Several flux tube-like structures appear in the SARI-dominated disk.}
    \label{fig:3D_many_modes}
\end{figure}

Figure~\ref{fig:3D_many_modes} shows 3D visualisations of the global disk with nearly azimuthal field above. A cut of the MRI disk is shown in the back, and a cut of the SARI disk in front. At least visually, the appearance of the non-axisymmetry in the SARI disk and field lines is reminiscent of what actual non-linear simulations obtain for a turbulent state \citep{matsumoto_tajima95, cylindrical_shearing_box}. Note that all these visualisations are in the linear phase, and hence the field remains dominant in the equilibrium direction. Further in the non-linear regime, true dynamo action is expected where the field will grow a strong radial component and align with the non-axisymmetry. However, the SARI field lines show more vertical excursions than the MRI field lines, which are predominantly displaced in the $r-\theta$ direction. Note the appearance of several flux tube-like structure in the SARI disk. These are the result of the opposite polarizations for the $\pm m$ SARIs. We checked this by considering the equivalent of Fig.~\ref{fig:SARI_islands} but for two values of $m$ with the same sign. We also checked that a collection of SARIs from the same $(m,k)$ wavevector produces not these complex twisted fields, but rather a coherent wave package of similar modes shifted radially. What is obvious from the comparison in Fig.~\ref{fig:3D_many_modes} is how the MRI superposition is always concentrated near the inner boundary, can never induce flux tube structures, and remains a mainly radial exchange without up-down phase speeds. The superposed SARIs are totally insensible to artificial radial boundaries, opposing $m$ modes can lead to apparent fine-scale flux tube structures and all modes have up or downwards phase speeds.

\subsection{$\alpha$-parameter}

As a basic consistency check to compare these linear visualisations to the vast literature on accretion disk instability and turbulence, we extract the averaged stresses from the SARI disks shown above. An important subject in shearing box and global simulations is the magnitude of Reynolds and Maxwell stresses resulting from MRI turbulence, which are the driver of angular momentum transport \citep{arlt_rudiger2001, bacchini2022, gorbunov24, bai_stone, begelman22, salvesen16}. We calculate a disk-averaged radial profile of the $\alpha$-parameter based on the Maxwell and Reynolds stresses as follows: 
\begin{equation} \label{eq:alpha}
    \alpha = \alpha_R + \alpha_M, \qquad \alpha_R = \langle \rho v_r v_\theta \rangle / \langle p \rangle, \qquad \alpha_M = \langle -B_r B_\theta \rangle / \langle p \rangle, \qquad \langle F(r) \rangle = \iiint f(r,\theta,z) d\theta dz / 2\pi 2\lambda_z
\end{equation}
where $p$ is the gas pressure, $\lambda_z$ the vertical wavelength, and the full field is taken into account for all quantities, i.e. $f_0 + f$. We average over $100$ gridpoints in both the $\theta$- and $z$-direction and then evaluate the stresses at the radius of the strongest growth, where they are maximal. Evaluating these non-linear quantities in the linear regime is a priori meaningful, since the values will be close to the linearised stresses. We quantify the stresses at the end of the linear phase, which is not uniform over the disk as the growth rates scale with $\Omega(r)$. Since the strongest-growing MRI dominates all other modes near the end of the linear regime and is closely attached to the inner boundary, it does not give a realistic view in terms of the stresses distributed over the disk. We hence exclude it from the present analysis and focus only on the non-axisymmetric SARI modes, which are localised in the interior of the disk.

\begin{figure}
    \centering
    \includegraphics[width=0.7\linewidth]{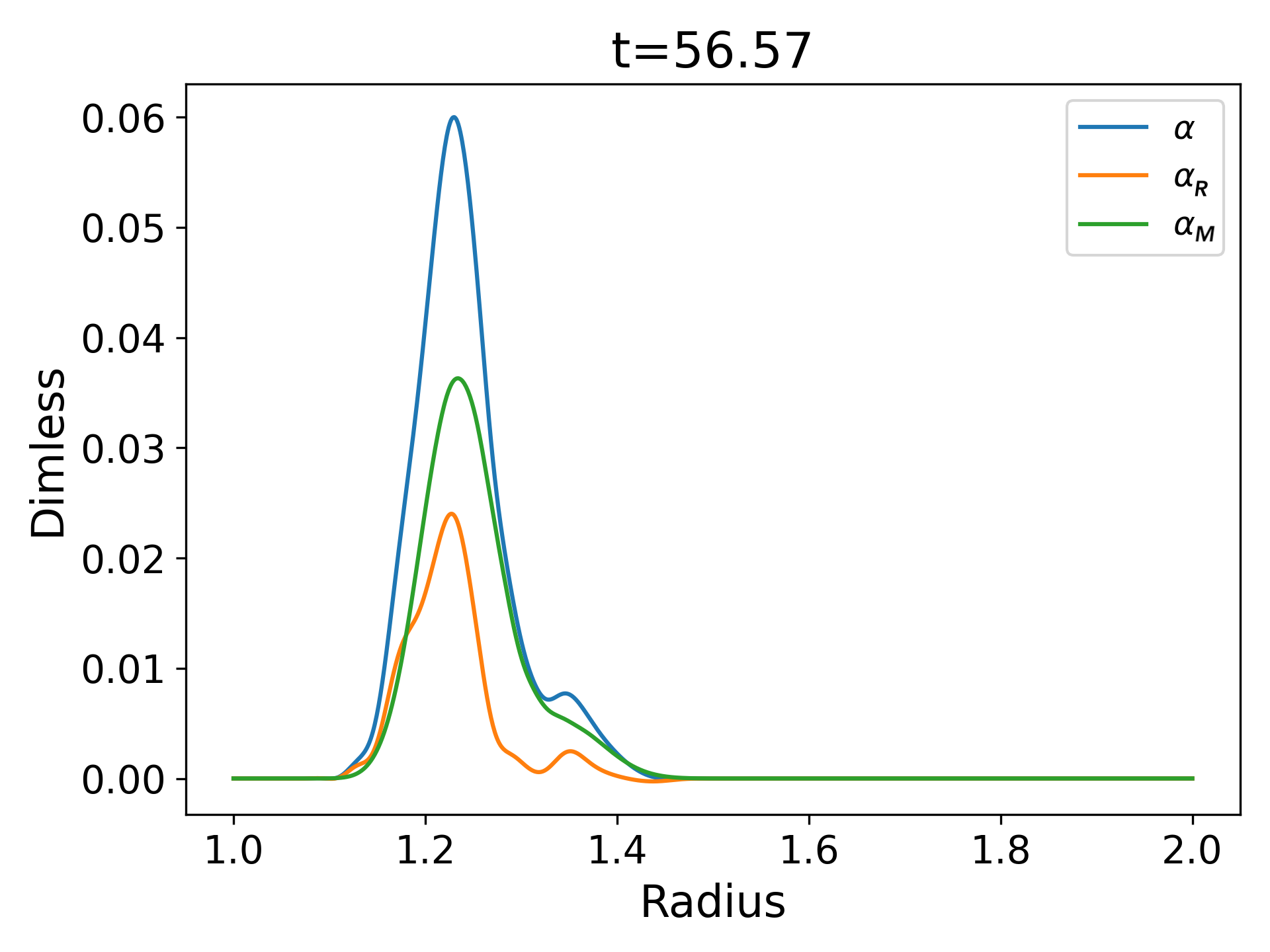}
    \caption{Disk-averaged radial profiles of stresses in the SARI with vertical field case at $t_\text{max}$. At this point, the fastest-growing modes near the inner parts of the disk dominate.}
    \label{fig:alpha_radially_averaged}
\end{figure}

It is well known from shearing box simulations that usually $\alpha_M > \alpha_R$, with $\alpha$ in the range $0.01 - 1.0$ after saturation of the MRI, depending on the initial conditions. Since the modes in this work are well in the linear regime, no saturation will occur. An example of the disk-averaged radial profiles of the stresses for a vertical background field is shown in Fig.~\ref{fig:alpha_radially_averaged}. It shows how at the end of the linear phase defined by the strongest-growing modes, the maximal stresses are limited to the region where those modes are located. The usual ordering $\alpha_M > \alpha_R$ is recovered. The maximal stresses at the end of the linear phase for all SARI disks are collected in Table \ref{tab:alpha} and reach orders of magnitude comparable to the known range of $\alpha$. The order of magnitude difference in stresses for fields that are more azimuthal versus more vertical aligns with the evidence that for a zero net flux field, $\alpha \sim 0.01$ \citep{Davis2010,simon2012}, whereas for a net vertical field, $\alpha \approx \beta^{-0.5}$ \citep{Hawley95,bai_stone,salvesen16,armitage_lecture_notes}, which in our case produces $\alpha \approx 0.31$. In non-linear simulations, before saturation, the time evolution of the $\alpha$ parameter is typically exponential over some rotation periods. The fact that our superposed linear instabilities already produce stresses at the lower limit of the saturated values, even though they enter the non-linear regime after only around one rotation period, implies that the linear dynamics may extend beyond the strictly defined linear regime. Answering the question of how much of the linear behaviour remains imprinted in the further non-linear evolution requires further understanding of the transition between the linear and non-linear growth (see also \citet{Latter2015}).

\begin{table}
	\centering
	\caption{Maximal values of the Shakura-Sunyaev $\alpha$ parameter and Reynolds and Maxwell stresses for the SARI datasets containing many superposed normal modes. The radius at which the stresses are maximal is also the radius where the fastest-growing modes are located.}
	\label{tab:alpha}
	\begin{tabular}{lllccc}
		\hline
		& $r_\text{max}$ & $\boldB_0$ & $\alpha$ & $\alpha_R$ & $\alpha_M$ \\
		\hline
        \textbf{SARI} & $1.23$ & Vertical & $5.99\times10^{-2}$ & $2.39\times10^{-3}$ & $3.61\times10^{-2}$ \\
        & $1.22$ & Helical & $8.38\times10^{-2}$ & $3.10\times10^{-2}$ & $5.27\times10^{-2}$ \\
        & $1.13$ & Azimuthal & $7.11\times10^{-3}$ & $2.46\times10^{-3}$ & $4.66\times10^{-3}$ \\
		\hline
	\end{tabular}
\end{table}

\section{Localised non-axisymmetric instabilities} \label{sec:local_vs_global}
In this final discussion section, we connect the present SARI analysis to some earlier works on localised non-axisymmetric instabilities in the local approximation.

\subsection{A continuum of unstable modes}
The idea of a dense 2D region of unstable modes in accretion disks with a toroidal background field was already proposed by \citet{terquem_papaloizou} (TP96), who themselves based the argument on a paper by \citep{Lin1993} regarding the Parker instability in vertically stratified shearing sheets. In these papers, an MHD operator $\mathcal{L}$ was considered containing the operators $\boldG$ and $U$ in a spectral equation equivalent to \eqref{eq:spectral_equation}. In particular, by assuming that $\omega$ satisfies a local eigenvalue equation like that given in \citet{blokland05} and then constructing appropriate trial functions $\boldxi(\boldk)$ in the limit of $|\boldk|\rightarrow\infty$, it was shown that such $\omega$ belong to the spectrum of $\mathcal{L}$ without assuming a Fourier transform in the radial and vertical directions (the wavenumbers just appear in the trial functions). Since the $\omega$ that follow from the local dispersion relation depend on both the local coordinates $(r_0,z_0)$ and the ratio of radial to vertical wavenumbers $k_r/k_z$, and are in general complex (i.e. overstable) with oscillation frequency in the Doppler range, it was then argued that a 2D region densely packed with unstable discrete modes or a 2D unstable continuous spectrum was obtained: indeed, for every $(r_0,z_0)$ and every value of $k_r/k_z$, there is an unstable mode in this sense, according to the proof in TP96. 

The SARI analysis also delivers a 2D continuum of unstable modes, but it differs in the following aspects: (1) the problem is Fourier analysed in the azimuthal and vertical directions, whereas the radial direction has a non-trivial variation; (2) the modes found are almost normal quasi-modes, and hence not exactly discrete or continuous. A description in terms of the tiny driving energy needed to maintain their time variation is available. A 2D region of unstable modes is then obtained from the global eigenvalue problem for a given value of $k_z$. For each frequency, the eigenfunctions are calculated and show a non-trivial radial variation where the radial wavenumber is not constant but varies over the eigenfunction. Similarly to the analysis of TP96, these modes are localised around a corotation radius $r_0$. However, the ratio $k_r/k_z$ plays no role of significance, as such an unstable 2D region is obtained for a fixed $k_z$. Hence, $k_r$ can take arbitrarily large values, as it was shown in GK22 that for growth rates $\nu\rightarrow0$ the eigenfunctions are oscillating extremely fast as they approach the continuum frequencies at marginal stability. In conclusion, the 2D continuum obtained by TP96 in the local limit for a toroidal field contains unstable modes located at various radii throughout the disk and with varying localisation radially and vertically, just like the SARIs, but these are true discrete or continuum modes instead of quasi-modes. Furthermore, quasi-continuum SARIs appear for all background field orientations.

\subsection{Local vs global in terms of wavenumbers}
TP96 then apply an initial value approach to the global, cylindrical problem for a variety of disk set-ups with or without vertical stratification. The resulting mode structure is very reminiscent of SARI wave packages, localised around a certain radius and radially oscillating. Growth occurs indefinitely, where the apparent wavenumbers $k_r$ and $k_z$ are ever-increasing until the grid resolution is reached. Following the analysis of \citet{ogilvie_pringle} and \citet{BKG24}, this is a particular feature of a purely toroidal background field, where the most unstable mode has arbitrarily large $k_z$. Further study is needed to determine whether the increasingly localised structure of the initial-value solutions is a form of transient growth at lower wavenumbers, or the result of an ever-increasing sequence of modes growing faster than those at lower $k_z$. 

This observation is related to the common assumption in shearing sheets that non-axisymmetric structure gets sheared away and that the radial wavenumber is hence time-dependent \citep{BH92b}. TP96, and a similar analysis by \citet{foglizzo_tagger95,foglizzo_tagger96}, who consider a shearing sheet with vertical stratification and a toroidal background field, found that this temporal dependence of $k_r$ can cause an instability to gradually change character in the local limit. For $k_r \gg k_z$, the modes are rather like the Parker instability, with mostly vertical displacements driven by buoyancy and a Schwarzschild-like criterion for instability, whereas for $k_r \ll k_z$, the instability acts more like a generalisation of the MRI in a toroidal field. 

Using Eq.~\ref{eq:fourier_polar}, we can determine a radial `wavenumber' for the SARI modes used in this work. Around corotation, we found for all modes that $k_r$ is in the range $1.75 - 2$ $k_z$ (and checked that the radial wavelength hence derived indeed matches that of the eigenfunctions). This places these global results in direct contradiction with those obtained in the local limit. The problem is only made worse by observing that, as mentioned before, $k_r\rightarrow\infty$ as $\nu\rightarrow0$ for fixed $k_z$ around marginal stability in the global SARI eigenvalue problem, and the analytical results of GK22 are best applicable to this regime. This discrepancy between the local and global approach warrants further investigation, but confirms our findings in \citet{BKG24} that a naive application of a WKB dispersion relation fails for the SARIs.

\section{Conclusion} \label{sec:conclusion}

This work aimed to visually explore the MRI and SARI in global accretion disks. We first established the spiral form of non-axisymmetric SARIs and their corresponding phase shifts. 

We then recovered the MRI mechanism of angular momentum exchange between fluid elements from the field line and eigenfunction perturbations, and generalised this picture (1) to more general background equilibrium magnetic field directions and (2) to the SARIs as well. Most importantly, we recovered that the MRI produces mostly radial displacement of the field lines, whereas the SARIs have an important vertical component as well. Another implication of the SARI phase shifts is that oppositely polarized SARIs can produce complicated field structures that seem prone to reconnecting. It would be interesting to explore the effect of finite resistivity on such modes, for which our current \texttt{Legolas} code can readily be used \citep{jordi_tearing24}. The inclusion of other non-ideal effects will no doubt greatly increase the complexity of the spectra, but also significantly broaden the scope for the study of non-axisymmetric modes in e.g. protoplanetary disks or the outer regions of accretion disks, where it may be relevant to include for example heating and cooling effects \citep{joris24}. Examples of similarly localised non-axisymmetric instabilities in viscous set-ups are discussed in \citet{ogilvie_proctor}. A recent study of global non-axisymmetric instabilities at low $m$ and $k$ in non-ideal disks can be found in \citet{haywood_ebrahimi}.

We then superposed many MRI and SARI modes to show how with linear instabilities only, a more realistic view of a `turbulent' accretion state can already be obtained by allowing for non-axisymmetric modes. With localised, transient non-axisymmetric instabilities continuously appearing on a turbulent `background', such a view may in fact not be too far off from a realistic non-linear evolution. In particular, the non-axisymmetric conditions required for dynamo action would only follow after non-linear evolution in a MRI disk, but appear naturally for the SARIs in the linear stage of the evolution. The question remains how much of a footprint the linear instabilities have on the ensuing non-linear evolution. The linear channel mode in local shearing boxes, for example, is at the same time also a solution to the non-linear evolution equations \citep{goodmanxu94}. \citet{Latter2015} showed that this does not necessarily generalise to global modes, since only global solutions at extremely large $k$ are approximate non-linear solutions. In their particular application with a radially constant vertical background field, they found that a global MRI with $k=9000$ maintained its exponential growth until $B_r / B_0 \approx 10$, well beyond the criterion \eqref{eq:criterion_linear} we used in this work. If such linear instability mechanisms leave an imprint on the ensuing non-linear evolution, it might be in the form of a mean field. Given recent light polarization data from the EHT project \citep{EHT} and theoretical efforts in linking polarization to structured and turbulent mean fields in disks \citep{barnier24}, it might one day be a possibility to observationally distinguish the mechanisms that are driving instability in accretion disks. 

A separate question is how a large collection of SARI modes would actually behave in a non-linear simulation. Because they are normal modes up to a tiny excitation energy, an initial-value approach would probably not change our current picture very much. In fact, overlaying many quasi-continuum SARI modes that belong to the same wavevector $(m,k)$ visually produces not a hugely complicated structure, but rather one larger, coherent wave package. In a non-linear simulation, however, these modes might influence each other in a non-trivial way. A non-modal analysis could shed light on whether the growth produced by SARIs, which typically have lower growth rates than the modes attached to the walls in a global model, can be non-exponential and perhaps faster than the typically assumed MRI \citep{squire_bhattacharjee}. It will be key to establish (1) the as yet unexplored connection between the quasi-continua and pseudomodes/non-modal growth \citep{squire_bhattacharjee}; and (2) whether such localised quasi-continuum SARI wave packages have a shearing box analog a la \citet{matsumoto_tajima95} obtained as a limit from the global formulation. In particular, whether non-axisymmetric growth observed in shearing box simulations originates from parasitic instabilities (especially tearing-like instabilities) that thrive on channel mode solutions \citep{goodmanxu94,latter09,miravet-tenes_parasitic} or truly other, perhaps transient, instabilities like the SARI taking over from the dominant MRI growth is an important related question.

\section*{Acknowledgements}

We thank Henrik Latter, Gordon Ogilvie, Mattias Brynjell-Rahkola and Hans Goedbloed for useful discussions and advice. NB would like to thank Dion Donné for assistance creating the Pyvista setup \citep{pyvista}, which was used to create the 3D visualisations in this work. The other visualisations were made using the Pylbo post-processing tool in \texttt{Legolas}, and Matplotlib \citep{matplotlib}. NB is supported by FWO Flanders fellowship 11J2622N. RK acknowledges funding from FWO project Helioskill G0B9923N, and from KU Leuven C1 project C16/24/010 (UnderRadioSun).

%%%%%%%%%%%%%%%%%%%%%%%%%%%%%%%%%%%%%%%%%%%%%%%%%%
\section*{Data Availability}

The data generated for the large collection of modes in Figs.~9 and 10 can be provided upon request to the corresponding author.


\bibliographystyle{mnras}
\bibliography{refs.bib}

%%%%%%%%%%%%%%%%%%%%%%%%%%%%%%%%%%%%%%%%%%%%%%%%%%

%%%%%%%%%%%%%%%%% APPENDICES %%%%%%%%%%%%%%%%%%%%%

\appendix

\section{Analytic expressions for MRI and SARI polarization} \label{sec:efs_analytical}

Only approximate statements can be made about the eigenfunctions described in Sec.~\ref{sec:efs_numerical}. The theory described in \citet{GKP19} expresses various MHD quantities in terms of the radial Lagrangian displacement $\chi := r\xi_r$ and $\chi'$. Assuming incompressibility and a thin, weakly magnetised disk with vertically localised modes (the approximations mentioned in Sec.~\ref{sec:equilibrium}), we derive explicit expressions for the perturbed velocity $\boldv$ given in Eq.~\eqref{eq:velocity_xi}, and the perturbed field $\boldB$ and show how the polarization properties from Sec.~\ref{sec:efs_numerical} follow.

First, we establish the Lagrangian displacement components $\xi_\theta$ and $\xi_z$ in terms of $\chi := \chi_r + i\chi_i$, based on their field-projected equivalents in 13.93 and 94 of \citet{GKP19}. Note that incompressibility can be used to determine one from the other, since
\begin{equation} \label{eq:incompressible_xi}
    \nabla\cdot \bxi = 0 \qquad \Leftrightarrow \qquad \frac{1}{r}\chi' + \frac{im}{r}\xi_\theta + ik\xi_z = 0. 
\end{equation}
Under the incompressible approximation, together with the thin disk assumptions $B_0^2 \ll 1$, $B_0^2\beta \gg 1$, $m^2 \ll k^2r^2$, we obtain
\begin{align} \label{eq:xi_components}
    \xi_\theta &= i \frac{m}{r^2k^2} \chi' - i \frac{2\omegat\Omega}{r(\omegat^2-\omega_A^2)}\chi, \\
    \xi_z &= i \frac{1}{rk}\chi' + i \frac{2m\omegat\Omega}{kr^2(\omegat^2-\omega_A^2)}\chi,
\end{align}
which is valid for both MRI (where $m=0$ simplifies the expressions) and SARI. Further, assuming that $\omegat \approx i\nu$, which is valid for the MRI because the Coriolis shift of the eigenfrequencies is small, and for the SARI if we assume that $r$ is around corotation so that $\omega_r - m\Omega(r) \approx 0$, we obtain the following expressions for the perturbed velocity based on Eq.~\eqref{eq:velocity_xi}:
\begin{equation} \label{eq:velocity_xi_explicit}
    \boldv = \nu \frac{\chi}{r} \mathbf{\hat{r}} + \left[ \frac{m\nu}{r^2k^2}i\chi' - \frac{2\nu^2\Omega + r\Omega'(\nu^2+\omega_A^2)}{r(\nu^2+\omega_A^2)}\chi \right]\boldsymbol{\hat{\theta}} + \left[ \frac{\nu}{rk}i\chi' + \frac{2m\nu^2\Omega}{kr^2(\nu^2+\omega_A^2)}\chi \right] \mathbf{\hat{z}},
\end{equation}
so that the corresponding real vector is given by 
\begin{equation} \label{eq:velocity_xi_explicit_real}
    \real(\boldv) = \nu \frac{\chi_r}{r} \mathbf{\hat{r}} + \left[ -\frac{m\nu}{r^2k^2}\chi_i' + \frac{(1.5\omega_A^2-0.5\nu^2)\Omega}{r(\nu^2+\omega_A^2)}\chi_r \right]\boldsymbol{\hat{\theta}} - \frac{\nu}{rk} \left[ \chi_i' - \frac{2m\nu\Omega}{kr^2(\nu^2+\omega_A^2)}\chi_r \right] \mathbf{\hat{z}},
\end{equation}
where we used that the rotation profile is Keplerian.

Finally, the magnetic field perturbation can be obtained through 
\begin{equation} \label{eq:field_xi_implicit}
    \boldB = \boldB_0\cdot\nabla\boldxi - \boldB_0\nabla\cdot\boldxi - \boldxi\cdot\nabla \boldB_0,
\end{equation}
which reduces under incompressible conditions to
\begin{equation} \label{eq:field_xi}
    \boldB = i\frac{F}{r} \chi \mathbf{\hat{r}} 
                + \left[ -\frac{mB_{z0}}{r^2k^2} \chi' - \left(\left(\frac{B_{\theta0}}{r}\right)' + i \frac{2F\nu\Omega}{r(\nu^2+\omega_A^2)}\right)\chi \right] \boldsymbol{\hat{\theta}} 
                + \left[ -\frac{F}{rk}\chi' - \frac{1}{r} \left((B_{z0})' - i \frac{2mF\nu\Omega}{kr(\nu^2+\omega_A^2)}\right)\chi \right] \mathbf{\hat{z}}, 
\end{equation}
where we define $F := mB_{\theta0}/r + kB_{z0}$. The real vector is then given by
\begin{equation} \label{eq:field_xi_real}
    \real(\boldB) = -\frac{F}{r} \chi_i \mathbf{\hat{r}} 
                + \left[ -\frac{mB_{z0}}{r^2k^2} \chi_r' - \left(\frac{B_{\theta0}}{r}\right)'\chi_r + \frac{2F\nu\Omega}{r(\nu^2+\omega_A^2)}\chi_i \right] \boldsymbol{\hat{\theta}} 
                + \left[ -\frac{F}{rk}\chi_r' - \frac{(B_{z0})'}{r}\chi_r  - \frac{2mF\nu\Omega}{kr^2(\nu^2+\omega_A^2)}\chi_i \right] \mathbf{\hat{z}}. 
\end{equation}

To establish the polarization properties of the MRI and SARI, we will apply a WKB-like argument around corotation $r_*$, assuming the eigenfunctions are strongly oscillating. In particular, this applies to the radially-varying part of the MRI modes and for the SARIs around corotation as $\nu \ll 1$. We furthermore use that the instabilities have to obey the instability criterion  $0<|\omega_A|<\sqrt{3}\Omega$, with strongest growth at $\omega_A = \sqrt{\frac{15}{16}}\Omega \approx \Omega \Rightarrow F \approx \Omega$, and that the maximum growth rate is $\nu \leq 0.75 \Omega$.

We also need to establish some empirical results regarding the eigenfunctions $\chi$. We have already mentioned that the MRI eigenfunction $\chi$ is approximately real and that the SARI eigenfunctions are truly complex. The reason for this is that the MRI has only a small Doppler-Coriolis frequency shift $\omega_r \sim 10^{-3}$ (GK22) (for truly compressible cases, this can become much larger, see \citet{das18}). Hence, $\omega \approx i\nu$ for the MRI, and as a result the basic second-order differential equation describing the radial displacement $\chi$ is approximately real (Eq.~(23) of GK22). We noted in Sec.~\ref{sec:efs_numerical} that this implies that $|\chi|$ has true zeroes for the MRI, in the sense that the zeroes of $\chi_{r,i}$ coincide with the zeroes of $\chi_{i,r}$. In the strongly oscillating parts of the eigenmodes, these zeroes approximately coincide with the extrema of the total pressure perturbation $\Pi$, since under the current approximations
\begin{equation} \label{eq:total_pressure}
    \Pi \approx -\frac{r\rho(\nu^2+\omega_A^2)}{k^2} \chi' + i\frac{2m\rho\nu\Omega}{k^2r^2} \chi.
\end{equation}
This confirms our findings about the radially-varying part of the MRI eigenfunctions in Sec.~\ref{sec:efs_numerical}, and also \citet{Latter2015}.

With this information, we can explain the polarization properties of the MRI from Sec.~\ref{sec:efs_numerical} from Eqs.~\eqref{eq:velocity_xi_explicit_real} and \eqref{eq:field_xi_real}, which are seen to be greatly simplified by taking $m=0$. If $\chi$ is a real function, then $v_z$ is trivially zero. Similarly, if $\chi$ is purely imaginary (after complex rotation), then $B_z$ is zero. Let us hence apply a general complex rotation $\chi e^{i\Phi}$ so that the eigenfunction is complex (but trivially so) and we can consider both $\chi_r$ and $\chi_i$. This corresponds to considering the eigenfunction at different $\theta$ and $z$, and hence different phase $\Phi$. Then the polarization of $\boldv$ follows from 
\begin{equation} \label{eq:polarization_MRI_v}
    v_r(r_*) \text{ maximal} \Leftrightarrow \chi_r(r_*) \text{ maximal} \Leftrightarrow v_\theta(r_*) \text{ maximal} \Leftrightarrow v_z(r_*)=0,
\end{equation}
since $\omega_A > \nu$ and $\chi_i'(r_*) = 0 = \chi_r'(r_*)$. Hence, $v_r$ and $v_\theta$ are in phase and $v_z$ is out-of-phase. A similar argument shows that the extrema of $v_z$ coincide with the zeroes of $v_{r,\theta}$. For the magnetic field polarization, we have that 
\begin{equation} \label{eq:polarization_MRI_B}
    B_r(r_*) \text{ maximal} \Leftrightarrow \chi_i(r_*) \text{ minimal} \Leftrightarrow B_\theta(r_*) \text{ minimal} \Leftrightarrow B_z(r_*)=0,
\end{equation}
since the terms involving $(B_{\theta,z})'$ are $O(B_0)$ whereas the other terms are $O(1)$. As a result, $B_{r,\theta}$ are in anti-phase and $B_z$ is out-of-phase. Furthermore, the zeroes of the components of $\boldv$ and $\boldB$ coincide.

For the SARI eigenfunctions, which are complex functions, $|\chi|$ is a strictly positive function describing the envelope of the `wave package' structure of the eigenfunctions. As a result, the zeroes of $\chi_r$ and $\chi_i$ have to interlace. If the eigenfunction oscillates fast enough (assuming small $\nu$), we can approximately say that, the extrema of $\chi_{r,i}$ coincide with the zeroes of $\chi_{i,r}$, so $\chi_{r,i}'(r_*)=0 \Leftrightarrow \chi_{i,r}(r_*)=0$. Furthermore, because the spirals are trailing as discussed in Sec.~\ref{sec:phase_speeds}, we can make the following association around the extrema of the eigenfunctions for instabilities:
\begin{align}
    \text{sgn}\left(\chi_r'(r_*)\right)& = -\text{sgn}(m)\text{sgn}\left(\chi_i(r_*)\right), \\\text{sgn}\left(\chi_i'(r_*)\right)& = \text{sgn}(m)\text{sgn}\left(\chi_r(r_*)\right).
\end{align}
For the damped equivalents of the SARIs, the signs are opposite because the spirals are leading in that case.

To explain the polarization of the SARIs, we first look at Eq.~\eqref{eq:field_xi_real}. Then we can make the following associations:
\begin{equation} \label{eq:polarization_SARI_B}
    B_r(r_*) \text{ maximal} \Leftrightarrow \chi_i(r_*) \text{ minimal},\quad \chi_r(r_*)=0 \Leftrightarrow B_\theta(r_*) \text{ minimal} \Leftrightarrow B_z(r_*) \text{ extremum}, \text{sgn}(B_z(r_*))=\text{sgn}(m),
\end{equation}
hence confirming that $B_{r,\theta}$ are in anti-phase and $B_z$ either in phase or anti-phase, depending on the sign of $m$. Here, we need to establish that the first term of $B_\theta$ is dominated by the third term, and that the first term of $B_z$ dominates the third term. Because of the strongly oscillating nature of the eigenfunctions, we generally find that $\chi' \gg \chi$. But for $B_\theta$, the factor of $\chi_r'$ is $\frac{mB_{z0}}{r^2k^2} \ll 1$ whereas the factor of $\chi_i$ is $O(1)$. For $B_z$, the coefficient of $\chi_r'$ is $O(1)$ but the coefficient of $\chi_i$ is smaller than $O(1)$ (due to the factor $m/k$). Hence, the term with the derivative dominates. 
% These assumptions are mutually consistent:
% \begin{equation} \label{eq:consistent_assumptions}

% \end{equation}
Similarly, the fact that $v_r$ and $v_\theta$ are in phase follows from a reasoning like that for the MRI in Eq.~\eqref{eq:polarization_MRI_v}, again using that the second term of $v_\theta$ dominates. For $v_z$, the first term dominates and has the same sign as $m$ when $v_{r,\theta}$ is maximal, again reproducing the polarization of Table~\ref{tab:polarizations}. The observation that the zeroes of $\boldv$ and $\boldB$ interlace now follows from the fact that the zeroes of $\boldv$ are determined by $\chi_r(r_*)=0 \Leftrightarrow \chi_i'(r_*)=0$, whereas the zeroes of $\boldB$ are determined by $\chi_i(r_*)=0 \Leftrightarrow \chi_r'(r_*)=0$. Hence, $\boldv \cdot \boldB \approx 0$. 

Finally, the polarization for the damped counterparts of the MRI/SARI in Table~\ref{tab:polarizations} is also nicely reproduced by replacing $\nu \rightarrow -\nu$ in the above equations. Some vector components flip sign because they depend on $\nu$, others keep the same sign because of their dependence on $\nu^2$, in accordance with our numerical observations.

%%%%%%%%%%%%%%%%%%%%%%%%%%%%%%%%%%%%%%%%%%%%%%%%%%

% Don't change these lines
\bsp	% typesetting comment
\label{lastpage}
\end{document}